\documentclass[reprint,superscriptaddress,amsmath,amssymb,aps,tightenlines
]{revtex4-2}
\usepackage{amsthm}
\usepackage{graphicx}
\usepackage{dcolumn}
\usepackage{color}
\usepackage{bm}
\usepackage{CJK}
\usepackage[ruled,linesnumbered]{algorithm2e}
\usepackage{hyperref}
\usepackage[math]{cellspace}
\newcommand{\de}[2]{\frac{\textrm{d} #1}{\textrm{d} #2}}

\newcommand{\bs}{\boldsymbol}

\newcommand{\mc}{\mathcal}

\begin{document}
\title{Weaving classical turbulence with quantum skeleton}
\author{Weiyu Shen}
\affiliation{State Key Laboratory for Turbulence and Complex Systems, College of Engineering, Peking University, Beijing 100871, PR China}
\author{Jie Yao}
\affiliation{Advanced Research Institute of Multidisciplinary Sciences, Beijing Institute of Technology, Beijing 100081, PR China}
\author{Yue Yang}
\email{yyg@pku.edu.cn}
\affiliation{State Key Laboratory for Turbulence and Complex Systems, College of Engineering, Peking University, Beijing 100871, PR China}
\affiliation{HEDPS-CAPT, Peking University, Beijing 100871, PR China}

\begin{abstract}
  Matter entanglement is a common chaotic structure in both quantum and classical systems. Turbulence can be pictured as a tangle of vortex filaments in superfluids and viscous vortices in classical fluids. However, it is hard to explain how the statistical properties of turbulence arise from elemental structures. Here we use the quantum vortex tangle as a skeleton to generate an instantaneous classical turbulent field with intertwined vortex tubes. Combining the quantum skeleton and tunable vortex thickness makes the synthetic turbulence satisfy key statistical laws and provides valuable insights for elucidating energy cascade and extreme events. By manipulating the elemental structures, we customize turbulence with desired statistical features. This bottom-up approach of ``weaving'' turbulence provides a testbed for analyzing and modeling turbulence.
\end{abstract}
\maketitle


Turbulence is recognized as one of the most complex physical systems~\cite{Sreenivasan1990Turbulence}.
Delving into the underlying structure is the foundation for understanding and regulating complex systems. 
Although bottom-up approaches have achieved significant success in physics~\cite{Coulais2016Combinatorial}, chemistry~\cite{Ding2023Polymer}, and biology~\cite{Henderson1975Three,Guido2006A},
it remains very challenging to decompose a turbulent field into elementary structures and explain the statistical properties of turbulence that arise from the superposition of simple flow fields.

Vortices are the key components of turbulence~\cite{Cardesa2017The}, constantly accommodating and releasing stress within the fluid, and shaping its behavior, akin to the sinews and muscles of a living organism~\cite{Küchemann1965Report,Moffatt1994Stretched}.
They are not randomly distributed in turbulence, but rather form coherent patterns~\cite{Hussain1986Coherent}, such as hairpin vortices~\cite{Adrian2007Hairpin} near walls, reflecting some underlying order beneath the seemingly random fluid motion.
Various structural-based vortex models, with spherical~\cite{Synge1943On}, tubular~\cite{Townsend1951On}, sheet~\cite{Corrsin1962Turbulent}, or spiral~\cite{Lundgren1982Strained} shapes, have been suggested to represent fine-scale structures of turbulence~\cite{Pullin1998Vortex}. 
On the other hand, massive observations of turbulence~\cite{She1990Intermittent,Jiménez1993The,Ishihara2013Thin, Matsuzawa2023Creation} reveal that the intense vortices are not a simple accumulation of elemental structures; the chaotic vortices, they always entangle with each other, forming a complex network~\cite{Xiong2019Identifying}.
Therefore, identifying simple universal structures that can capture essential features and dynamics of turbulence is a formidable challenge.

An inverse approach to understanding turbulence assembles a turbulent field from the bottom-up -- using fine structures as building blocks.
By comparing synthetic and real turbulence, we can distinguish the effects of different elemental structures or their combinations on turbulence statistics. 
Existing synthetic methods of turbulence, such as Fourier modes~\cite{Fung1992Kinematic}, linear-eddy models~\cite{Kerstein1988A}, fractal models~\cite{Juneja1994Synthetic}, and multiscale Lagrangian map~\cite{Rosales2006A} can generate random velocity fields with some coherence. Still, they struggle to simultaneously reproduce fine-scale vortices, turbulence statistics, and intermittency in 3D turbulence (table S1).

To shed light on the elemental structure of turbulence, we develop a novel bottom-up method to ``weave'' an instantaneous turbulent field using intertwined vortices with customizable fine-scale features.
This provides a unique way to generate turbulent fields for assessing turbulence theory and developing structure-based models. 
It eliminates the need for simulating flow evolution and can be employed as tailored flow data for further turbulence simulations and machine learning training. 

\begin{figure}
	\centering
	\includegraphics[width=0.9\columnwidth]{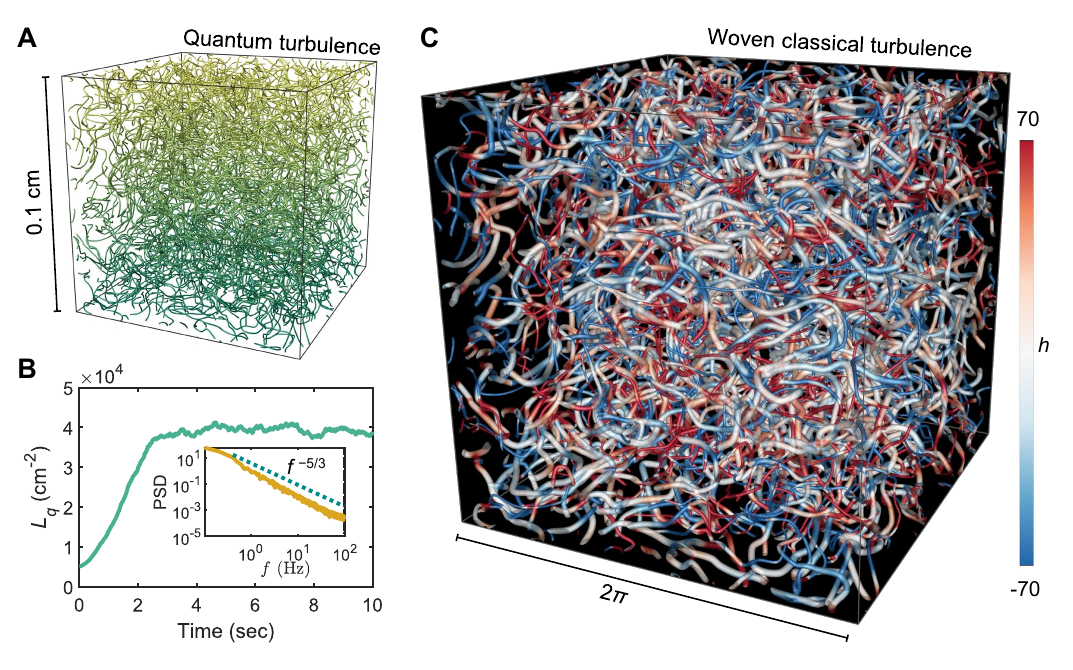}
	\caption{\textbf{Constructing a turbulent field in classical viscous flow with vortex filaments of quantum turbulence.}
    (\textbf{A}) A tangle of vortex filaments in the statistically stationary stage of quantum turbulence obtained from the simulation of superfluid helium II using the vortex filament method. The simulation is implemented in a cube of size $\mathcal{L} =0.1$ cm with periodic boundary conditions (movie S1). 
    (\textbf{B}) Temporal evolution of the quantum vortex-line density $L_q$, the vortex length per unit volume. After rapid growth at early times, the length of quantum vortex filaments gradually reaches a statistically stationary stage. The inset shows the frequency spectrum (power spectral density) of the fluctuations of $L_q$ during the stationary stage. The dashed line denotes the scaling of $f^{-5/3}$, where $f$ is the frequency.
    (\textbf{C}) Vortex-surface visualization of an instantaneous woven classical turbulent field with the quantum skeleton in (A).
    The skeleton size is non-dimensionalized into a periodic box of size $\mathcal{L}=2\pi$. Each viscous vortex tube is constructed with a quantum filament as the tube centerline, and its vorticity profile is a Gaussian function. The core size  $\sigma (s)=0.015 (1+0.5\sin s)$ of the vortex tube varies along the arc length parameter $s$. The isosurface is the normalized vortex-surface field~\cite{Yang2010On}  $\phi_v=0.1$ and color-coded by the helicity density $h = \boldsymbol{u} \cdot \boldsymbol{\omega}$, the dot product of the fluid velocity $ \boldsymbol{u} $ and the vorticity $ \boldsymbol{\omega}=\nabla \times \boldsymbol{u} $.}
	\label{fig:1}
\end{figure}

\textbf{Skeleton of turbulence.}
Considering the vortices should be chaotic and entangled, we propose that quantum turbulence~\cite{Barenghi2014Introduction} can serve as the ``skeleton'' for classical turbulence.
Superfluid turbulence, governed by quantum physics, manifests as a tangle of vortex filaments, exhibiting remarkable statistical similarities with classical turbulence \cite{Kivotides2003Quantized, Polanco2021Vortex}. 
We simulated the vortex tangle of superfluid helium II using the vortex filament method~\cite{Hanninen2014Vortex} in a cube of size $ \mathcal{L} =0.1$ cm with periodic boundary conditions (see movie~S1 and Supplementary materials, Materials and Methods~A).
After a rapid initial transition, the length of superfluid vortex filaments reaches a statistically stationary, homogeneous, and isotropic state (Figs.~\ref{fig:1}A and B).
In this stage, the frequency spectrum of the vortex-line density fluctuation obeys the $-5/3$ scaling, consistent with low-temperature experiments \cite{Bradley2008Fluctuations} (inset in Fig.~\ref{fig:1}B).
These quantum vortex filaments are used as centerlines, i.e., the skeleton, of intertwined viscous vortex tubes in weaving turbulence.

Subsequently, we generated the axisymmetric Burgers vortex tubes from the skeleton.
Here, the radial vorticity distribution is $\omega_{s}(\rho)=\Gamma\exp \left[-\rho^2/(2 \sigma^2)\right]/(2 \pi\sigma^2)$ with the radial distance $\rho$ from the centerline and the normalized circulation $\Gamma=1$. This simple exact solution of the Navier--Stokes equations is an attractive candidate for modeling fine-scale turbulence and encapsulates the stretching by the local strain and the dissipation by viscosity.
In particular, the vortex tube has a variable core size $\sigma(s)$ along arc-length $s$ of the centerline, which characterizes the evolution of vortices with multiscale features.

We have developed a general method to construct the vorticity for vortex tubes with arbitrary centerline topology, differential twist, and variable thickness (see Supplementary materials, Materials and Methods B, and Supplementary Text, Section II for details). Figure~\ref{fig:1}C shows an example of a constructed 3D turbulent flow field with the quantum skeleton (in Fig.~\ref{fig:1}A).
This ``woven turbulence'', a special synthetic turbulent field, consists of interwind, worm-like vortices. Statistically, it is homogeneous and isotropic, as evidenced by the Gaussian probability density functions (PDFs) of fluctuating velocity components (Fig.~S5).

\textbf{Flow statistics and energy cascade.} 
Turbulence statistics are estimated from the instantaneous woven field based on the celebrated Kolmogorov 1941 theory \cite{Kolmogorov1991The}, with measuring the turbulent kinetic energy $k_t=3.12$ and enstrophy $\left\langle \Omega \right\rangle =2.72\times10^3$ (See Supplementary Text, Section III~A for details).
We calculated the key statistics, the mean dissipation rate $\left\langle \epsilon \right\rangle=3.67 $, the kinetic viscosity $\nu=6.67\times10^{-4}$, and the Taylor Reynolds number $Re_\lambda=161.4$ (table~S2).

\begin{figure}
	\centering
	\includegraphics[width=0.9\columnwidth]{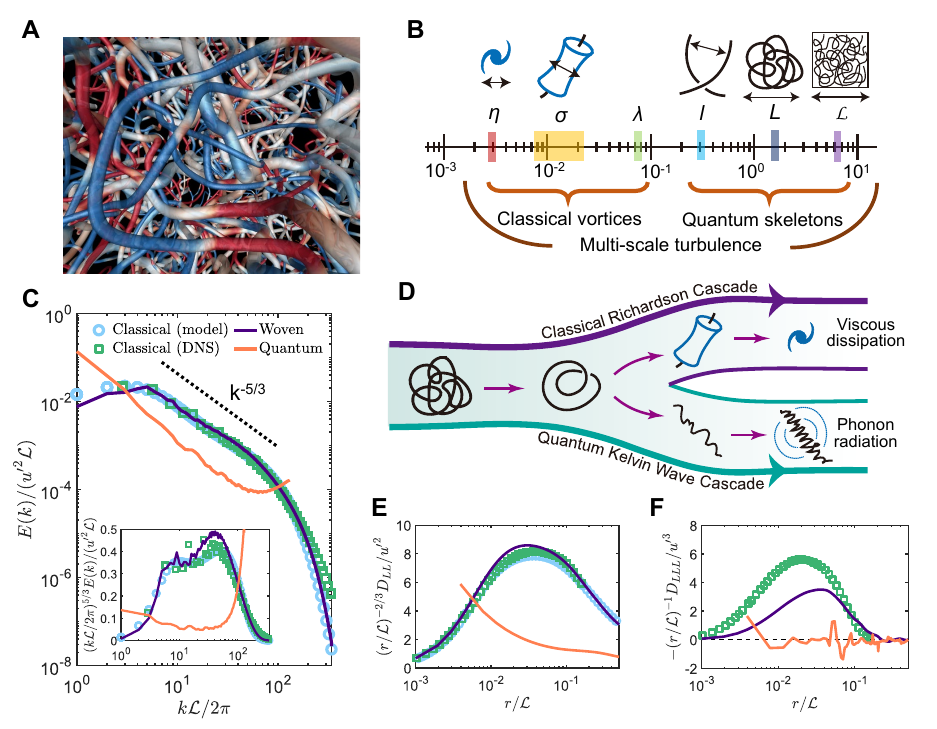}
	\caption{\textbf{Scales, cascade, and statistics of classical, woven, and quantum turbulence.}
    (\textbf{A}) Intertwined vortex tubes with different core sizes in woven turbulence. Local visualization is presented by the isosurface of the normalized vortex-surface field $\phi_v=0.1$ color-coded by the helicity density $h$.
    (\textbf{B}) Key length scales in multi-scale woven turbulence. The quantum skeleton frames the large scales including the average distance $l$ between vortices, integral length scale $L$, and box size $\mathcal{L}$, while classical vortices supply the small scales including the Kolmogorov length scale $\eta$, core size $\sigma$, and Taylor microscale $\lambda$. The specific values are calculated in table S2.
    (\textbf{C}) Rescaled 3D energy spectra of classical (symbols), woven (purple line), and quantum (orange line) turbulence. The dashed line shows the scaling of $k^{-5/3}$, where $k$ is the wavenumber in the Fourier space. Here, $u^\prime$ is turbulent root-mean-square velocity, and Taylor Reynolds number is $Re_\lambda=161.4$. The circles denote data calculated from the model \cite{Su2023Simple} ($Re_\lambda=161.4$) for classical turbulence (see Supplementary Text, Section III~C for more details). The squares denote data obtained from a direct numerical simulation ($Re_\lambda=161.4$) of classical turbulence (see Supplementary materials, Materials and Methods for more details). Inset plots the rescaled compensated energy spectra which clearly exhibit the inertial range and bottleneck effect. Data rescaled by Kolmogorov scales are supplemented in Fig.~S6.
    (\textbf{D}) Schematic of classical Richardson energy cascade and quantum Kelvin-wave energy cascade. The introduction of finite thickness of viscous vortex tubes into the turbulence with only quantum vortex filaments results in a shift from quantum cascade to classical one.
    (\textbf{E}) Rescaled longitudinal second-order velocity structure function $D_{LL}(r)=\left\langle\left[u_1\left(\boldsymbol{x}+\boldsymbol{e}_1 r\right)-u_1(\boldsymbol{x})\right]^2\right\rangle$ of the same data as in (C).
    (\textbf{F}) Rescaled longitudinal third-order velocity structure function $D_{LLL}(r)=\left\langle\left[u_1\left(\boldsymbol{x}+\boldsymbol{e}_1 r\right)-u_1(\boldsymbol{x})\right]^3\right\rangle$ of the same data as in (C). }
	\label{fig:2}
\end{figure}

The various structural features introduced in our bottom-up construction lead to a multi-scale nature of the woven turbulent flow field.
The vortex tubes are intertwined in chaotic patterns, with varying entanglement degrees and core sizes (Fig.~\ref{fig:2}A). These features affect the range of scales in turbulence.
All critical length scales are calculated and annotated in Fig.~\ref{fig:2}B.
The quantum skeleton encodes large-scale information including the integral length scale $L=k_t^{3/2}/\left\langle \varepsilon \right\rangle$ and average distance $l$ between vortices, which characterize the denseness of vortex tubes. 
The viscous vortices generated on the skeleton expand the range of scales from the zero measure of quantum filaments to the finite Kolmogorov scale $\eta=\left(\nu^3 / \varepsilon\right)^{1 / 4}$, which characterizes the viscous dissipation of kinetic energy.
By this means, the integration of the quantum skeleton and vortex tubes endows the woven turbulence with a wide range of length scales as in classical turbulence.

To explore the energy cascade~\cite{Yao2022Vortex} in the woven turbulence, figure~\ref{fig:2}C compares the rescaled 3D energy spectra of classical, woven, and quantum turbulence.
The energy spectrum of woven turbulence agrees well with the model result of classical turbulence \cite{Su2023Simple} and direct numerical simulation (DNS) result and satisfies the scaling of $k^{-5/3}$ in the inertial range.
The compensated energy spectrum in the inset of Fig.~\ref{fig:2}C reveals the bottleneck effect \cite{Lohse1995Bottleneck} between the inertial and dissipation ranges, which demonstrates that woven turbulence captures the cascade of turbulence at small scales.
By contrast, the spectrum of quantum turbulence only shows the scaling of $k^{-5/3}$, whereas it is distinct from that of classical and woven turbulence at small scales due to the lack of finite-sized tubular geometry.

The cascade picture elucidates why woven turbulence can recover classical turbulence statistics from tangled filaments by introducing viscous vortex tubes.
Figure~\ref{fig:2}D sketches the bifurcation of energy cascades in woven and quantum turbulence.
In quantum turbulence, the reconnection of vortex filaments generates high-frequency Kelvin waves, and the energy is dissipated into heat by phonon radiation at the atomic scale \cite{Vinen2001Decay}.
In woven turbulence, the quantum Kelvin-wave cascade is transformed into the classical Richardson cascade by supplementing the viscous vortex scales below the vortex filament spacing $l$, so that the energy in the inertial range can be transferred and then dissipated by viscosity at the Kolmogorov scale.
This cascade is supported by the spectral energy flux and transfer function in Fig.~S7.

Longitudinal velocity structure functions further confirm the consistent statistics in woven and classical turbulence.
In Fig.~\ref{fig:2}E, the second-order structure function $D_{LL}(r)$, rescaled by Kolmogorov's scaling of $r^{-2/3}$ in the inertial range, is consistent with that of classical homogeneous isotropic turbulence.
The woven turbulence also displays the negative third-order structure function $D_{LLL}(r)$ with the scaling of $r^{-1}$. This unique characteristics of classical turbulence is absent in Gaussian random fields and quantum turbulence.
Moreover, the higher-order structure functions of woven turbulence have the anomalous scalings in classical turbulence (Fig.~S6).

\begin{figure}
	\centering
	\includegraphics[width=\columnwidth]{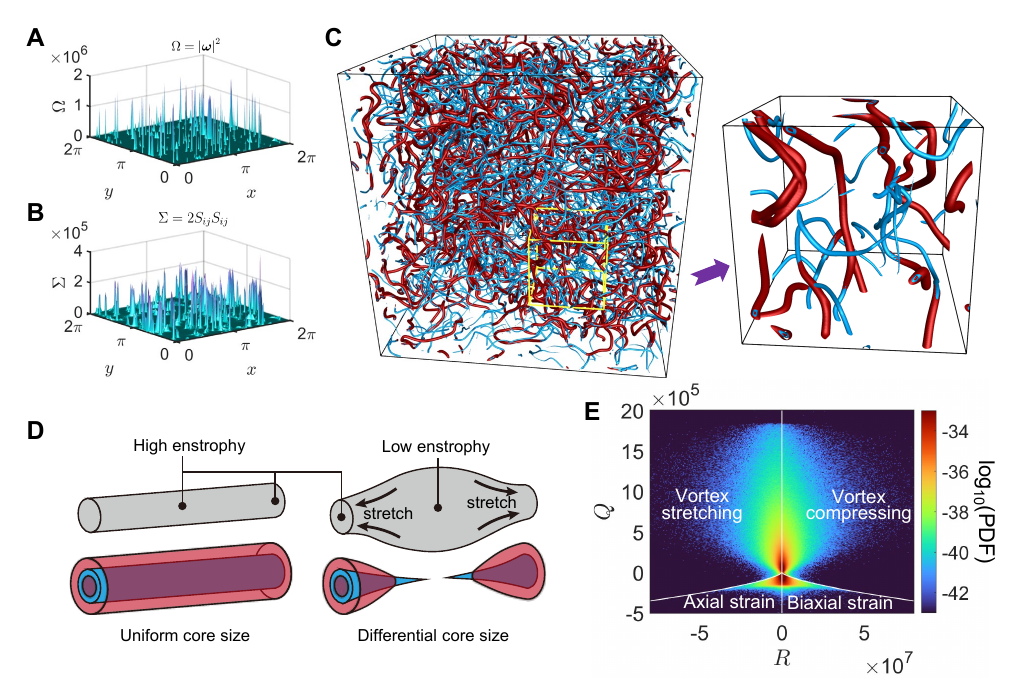}
	\caption{\textbf{Intermittency and extreme events in woven classical turbulence.} 
    (\textbf{A}) Intermittent distribution of the enstrophy $\Omega=|\boldsymbol{\omega}|^2$ in the plane at $z=0$, with the vorticity $\boldsymbol{\omega}$. 
    (\textbf{B}) Intermittent distribution of the dissipation divided by viscosity $\Sigma=\varepsilon/\nu=2S_{ij}S_{ij}$ in the plane at $z=0$, with the strain-rate tensor $S_{ij}$. 
    (\textbf{C}) Isosurfaces of the enstrophy (blue) and dissipation (red) $\Sigma=\Omega=10^5=18.4/\tau_K^2$, with $\left\langle \Sigma \right\rangle = \left\langle \Omega \right\rangle=1//\tau_K^2 $ and the Kolmogorov time scale $\tau_K=(\nu/\left\langle \varepsilon\right\rangle )^{1/2}$. Close-up view for the region is marked by the yellow box.
    (\textbf{D}) Schematic of the vortex surface (gray), enstrophy isosurface (blue), and dissipation isosurface (red) in vortex tubes with uniform and differential core sizes.
    (\textbf{E}) Joint PDF of the second and third invariants ($Q$ and $R$) of the velocity-gradient tensor. The red region marks the location where most extreme events are found. White lines denote $R = 0$ and Vieillefosse line $27R^2 + 4Q^3 = 0$, which divide the $R$-$Q$ plane into four regions: vortex stretching, vortex compressing, axial strain, and biaxial strain.}
	\label{fig:3}
\end{figure}

\textbf{Intermittency and extreme events.}
The coexistence of ordered and random natures causes classical turbulence to exhibit intermittency and extreme events, which are highly challenging to present in synthetic turbulence.
In woven turbulence, distributions of the local enstrophy $\Omega=|\boldsymbol{\omega}|^2$ and dissipation  (divided by viscosity) $\Sigma=\varepsilon/\nu=2S_{ij}S_{ij}$, measuring the strength of the velocity gradient tensor $S_{i j}=\partial u_i / \partial x_j$~\cite{Buaria2019Extreme}, are highly intermittent (Figs.~\ref{fig:3}A and B) as in classical turbulence.
In isotropic turbulence, their volume averages $\left\langle \Omega \right\rangle =\left\langle \Sigma \right\rangle$ are the same, but the distributions of their local structures are different.
Isosurfaces of $\Omega$ and $\Sigma$ with the same threshold occupy different parts of the vortex tube (Fig.~\ref{fig:3}C). Dissipation isosurface partly envelops the enstrophy one.
Our vortex model with varying core sizes explains that the mismatch between the enstrophy and dissipation is due to vortex stretching.
In Fig.~\ref{fig:3}D, these two quantities are co-located in a vortex tube with a constant core size, while variable core sizes lead to more intense dissipation being concentrated where the vortex is stretched.

The flow structure of extreme events can be classified and identified through $Q(\boldsymbol{x})=-\left(S_{i j} S_{j i}\right) / 2$ and $R(\boldsymbol{x})=-\operatorname{det}\left[S_{i j}\right]$, the second and third invariants of $S_{i j}$~\cite{Meneveau2011Lagrangian}.
The joint PDF of $R$ and $Q$ shows the distribution of the local topology of extreme events (Fig.~\ref{fig:3}E). The $R$-$Q$ plane is divided into four regions  (vortex stretching, vortex compressing, axial strain, and biaxial strain) by $R = 0$ and the Vieillefosse line $27R^2 + 4Q^3 = 0$~\cite{Moffatt2021Extreme}.
The extreme events are dominated by vortex stretching and compressing. However, unlike the classical teardrop shape, the $R$-$Q$ PDF in woven turbulence has a butterfly shape that lacks the fine-scale structural transition from vortices (upper part) to strains (lower part).
It is mainly attributed to our vortex model, which is compact in a tube region.
In contrast, real turbulence diffuses vorticity throughout the entire space, so that extreme events involve low-vorticity regions near $Q=0$.

Another notable difference is the symmetrical and asymmetrical $R$-$Q$ PDFs in woven and real turbulence, respectively.
We performed a DNS for decaying turbulence with the initial woven field, as described in Section~IV of the Supplementary Text.
When the vortex tubes start to interact with each other, well before the onset of fully developed turbulence, a part of the axial strains are rapidly converted into biaxial strains, resulting in an asymmetric $R$-$Q$ shape (Figs.~S9 and S10). 
Subsequently, small-scale vortices produced by the vortex interaction fill the gap between the vortex tubes, and the $R$-$Q$ shape evolves into a typical teardrop shape, implying that the local topology highly depends on the small-scale flow structures rather than the large-scale vortex skeleton.

\begin{figure}
	\centering
	\includegraphics[width=\columnwidth]{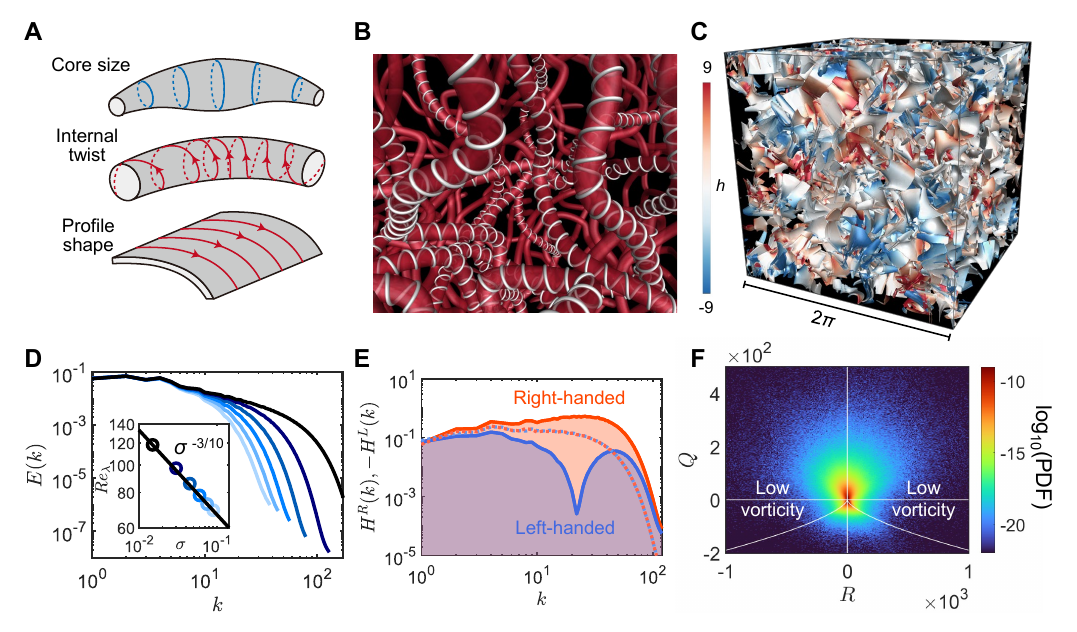}
	\caption{\textbf{Customization of turbulence.}
    (\textbf{A}) Schematic of three customizable features for an elemental structure of turbulence: core size, internal twist, and profile shape.
    Red directed lines represent typical vortex lines on the gray vortex surfaces.
    (\textbf{B}) Local visualization of woven classical turbulence with strong internal twist. Vortex lines with right-handed local twist rate $\eta_t=50$ are highly coiled within vortex tubes and the core size $\sigma=0.03$ is set to be uniform. Vortex tubes are visualized by the vortex-surface field $\phi_v=0.1$ and color-coded by the helicity density $h$. Some typical vortex lines (gray) are drawn on the vortex surfaces.
    (\textbf{C}) Visualization of turbulence synthesized by vortex sheets. The vorticity isosurface $|\boldsymbol{\omega}|=10$ is color-coded by $h$.
    (\textbf{D}) Energy spectra of woven turbulence with various core sizes from $\sigma=0.015$ (light) to $\sigma=0.09$ (dark), where $k$ is the wavenumber of in the Fourier space. The inset is the Taylor Reynolds number with various core sizes with the scaling of $\sigma^{-3/10}$ (solid line).
    (\textbf{E}) Helicity spectra for the right- ($H^R$, red) and left-handed ($H^L$, blue) polarized components of woven turbulence with internal twist in (B). Two dotted lines show the helicity spectra of woven turbulence without internal twist as a reference. Purely right- or left-handed polarized components of helicity are separated based on the helical wave decomposition (see Supplementary materials, Supplementary Text, Section III~E for details).
    (\textbf{F}) The joint PDF of the second and third invariants ($Q$ and $R$) of the velocity-gradient tensor of turbulence synthesized by vortex sheets in (C). White lines denote $R = 0$, $Q = 0$, and Vieillefosse line $27R^2 + 4Q^3 = 0$. The area near $Q=0$ represents a low vorticity region.}
	\label{fig:4}
\end{figure}

\textbf{Customizing turbulence.}
Besides unveiling insights into turbulent structures, the method of weaving turbulence also enables customizing a turbulent field with desired properties, such as the Reynolds number, turbulent kinetic energy, and helicity~\cite{Scheeler2017Complete,Meng2023Evolution}, by precisely controlling the geometry and topology of elemental vortical structures (see Fig.~\ref{fig:4}A for example). 
In particular, this method can generate turbulent fields that are valuable in theoretical study but difficult or costly to obtain through experiments and numerical simulations.

The quantum skeletons with more vortex filaments and higher entanglement correspond to larger turbulent kinetic energy and Reynolds number. 
A more valuable question is how the vortex geometry affects turbulence.
We manipulated the core size and internal twist~\cite{Shen2023Role} of vortex tubes (Fig.~\ref{fig:4}A) with the same quantum skeleton.
Increasing core size $\sigma$ can reduce the width of the inertial range of woven turbulence, and increase the Reynolds number with a power-law scaling of $Re_\lambda\sim\sigma^{-3/10}$ (Fig.~\ref{fig:4}D). 
In general, the scale $\sigma$ of vortex tubes is smaller than the scale $l$ of the quantum skeleton (see Fig.~\ref {fig:2}B). Increasing the core size of coherent vortices brings the scales of viscous vortices closer to large scales of the skeleton, leading to an overall contraction of length scales.

The internal twist of vortex lines within vortex tubes is a key topological component of helicity, altering the flow chirality \cite{Scheeler2017Complete}.
The local twist rate~\cite{Shen2023Role} $\eta_t$ of the vortex lines along a periodic vortex centerline $\mathcal{C}$ contributes to the twisting component $T_{w}=\oint_{\mathcal{C}} \eta_t/(2\pi) \mathrm{~d} s$ in the topological decomposition of total helicity~\cite{Scheeler2014Helicity,Shen2022Topological}
\begin{equation}
	\begin{aligned}
		H&\equiv\iiint \boldsymbol{u} \cdot \boldsymbol{\omega} \, \mathrm{d} \mathcal{V}\\
		&=\sum_{i \neq j} \Gamma_{i} \Gamma_{j} L_{k,ij}+\sum_{i} \Gamma_{i}^{2}(W_{r,i}+T_{w,i}),
	\end{aligned}
\end{equation}
where each index labels a single vortex tube, and the writhe $W_r$ and link $L_k$ are only related to the knotting and linking of the vortex skeleton.
In Fig.~\ref{fig:4}B, we constructed a helical turbulent field with highly coiled vortex lines as right-handed spirals inside the vortex tube.
Figure~\ref{fig:4}E shows the chirality from the helicity spectrum.
Right-handed twist waves attenuate the left-handed polarization of the helicity spectrum at a given wavenumber, resulting in a predominant right-handed chirality. In contrast, woven turbulence without internal twist has a balanced helical wave decomposition with $H\approx 0 $.

Modifying the cross-sectional shape of elemental vortices is effective in customizing turbulence statistics.
As an example, we use the vortex sheet, another candidate of the elemental structure, to weave a turbulent field in Fig.~\ref{fig:4}C.
The resulting energy spectrum exhibits the scaling law of $k^{-7/3}$ (Fig.~S12A), which is steeper than the $k^{-5/3}$ contributed by vortex tubes. 
This difference is consistent with the theoretical scaling of an ensemble of elemental vortices, i.e., $E\sim k^{-2}$ for sheet-like structures~\cite{Townsend1951On} is steeper than $E\sim k^{-1}$ for tube-like structures.
Moreover, the $R$-$Q$ joint PDF in the turbulence woven by vortex sheets has higher values near $Q=0$ than that woven by vortex tubes (Fig.~\ref{fig:4}F), because vortex sheets cover larger spatial extent than vortex tubes, resulting in increased regions with low vorticity and high strain rate in turbulence.


\textbf{Discussion.}
In summary, our bottom-up approach of ``weaving'' turbulence, utilizing adjustable fine structures as building blocks, offers a testbed for exploring and understanding elemental structures of turbulence.
By utilizing quantum vortex filaments as ``skeleton'', we have generated classical turbulent fields consisting of intertwined vortex tubes. The combination of the quantum skeleton and tunable tube thickness ensures that the woven turbulence adheres to key statistical laws, including the -5/3 scaling of the energy spectrum, negative third-order structure function, and positive spectral energy flux in the inertial range. By manipulating the elemental structures, we can tailor turbulent flow fields with different Reynolds numbers, helicities, and local flow geometries.

The customizable woven turbulence can be used to assess turbulence theory~\cite{Davidson2015Turbulence}, develop structure-based turbulence models~\cite{Pullin1998Vortex}, and offer a playground for studying multi-physics coupled flows~\cite{Aharon2022Direct,Dong2022Reconnection} and complex fluids~\cite{Wu2017Transition,Duclos2020Topological}. 
Although the present construction method may not capture all the intricacies of real turbulence, it opens up an avenue for exploring turbulence in the context of matter research and seeking connections between classical/quantum turbulence and other complex systems. 
Its low cost facilitates generating large training data for machine learning of turbulence~\cite{Novati2021Automating}.
By harnessing this powerful tool, one can incorporate diverse elemental structures, such as Lundgren's spiral vortex~\cite{Lundgren1982Strained}, into the woven flow field, and employ quantum wall flows~\cite{Baggaley2013Vortex} to weave wall turbulence with complex boundary conditions. 
This approach promises to illuminate valuable structures and their functionalities in understanding and modeling turbulence.


\begin{center}
	{\textbf{Acknowledgments}}
\end{center}
The authors are grateful to A. W. Baggaley for providing the code of quantum turbulence simulation, and thank C. F. Barenghi and S. Xiong for helpful comments.
Numerical constructions and simulations were carried out on the Tianhe-2A supercomputer in Guangzhou, China. 
This work has been supported by the National Natural Science Foundation of China (Grant Nos.~11925201 and 11988102), the National Key R\&D Program of China (No.~2020YFE0204200), and the Xplore Prize.

\begin{center}
	{\textbf{Author contributions}}
\end{center}
W.S. and Y.Y. conceived the idea and designed the research; Y.Y. supervised the project. W.S. constructed woven turbulence, simulated quantum turbulence, and processed data. J.Y. simulated classical turbulence. W.S. and Y.Y. wrote the original draft; all authors discussed the results and reviewed the final manuscript. All the authors have given approval for the manuscript.

\clearpage
\onecolumngrid

\clearpage

\clearpage

\clearpage


\setcounter{table}{0}
\setcounter{figure}{0}
\setcounter{section}{0}
\setcounter{equation}{0}
\renewcommand{\thetable}{S\arabic{table}}
\renewcommand{\thefigure}{S\arabic{figure}}
\newpage
\clearpage
\begin{center}
	\large
	\textbf{Weaving classical turbulence with quantum skeleton\\Supplementary material}\\
	\mbox{} \\
	\large
	Weiyu Shen\\
	\mbox{} \\
	{\it{State Key Laboratory for Turbulence and Complex Systems, College of Engineering,\\
	Peking University, Beijing 100871, PR China}}\\
	\mbox{} \\
	Jie Yao\\
	\mbox{} \\
	{\it{Advanced Research Institute of Multidisciplinary Sciences,\\
	Beijing Institute of Technology, Beijing 100081, PR China}}\\
	\mbox{} \\
	Yue Yang\\
	\mbox{} \\
	{\it{State Key Laboratory for Turbulence and Complex Systems, College of Engineering and HEDPS-CAPT\\
			Peking University, Beijing 100871, PR China}}\\
\end{center}

\newpage

\tableofcontents
\clearpage

\section*{Materials and Methods}
\subsection{Simulation of quantum turbulence}
We generate a tangle of vortex filaments in superfluid turbulence without the normal fluid using the vortex filament method (VFM) \cite{Hanninen2014Vortex}. 
The vortex filaments are further used as centerlines in constructing viscous vortex tubes with finite thickness.

In the VFM, the vortex filament moves according to the total induced velocity solved by the de-singularized Biot-Savart integral
\begin{equation}
	\frac{\mathrm{d}\boldsymbol{s}}{\mathrm{d}t}=\frac{\kappa_q}{4\pi} \boldsymbol{s}'\times\boldsymbol{s}''\ln\left(\frac{2\sqrt{l_+l_-}}{ \mathrm{e}^{1/2}a_0}\right) +\frac{\kappa_q}{4\pi}\int_{\boldsymbol{s}_1\neq\boldsymbol{s}}\frac{(\boldsymbol{s}_1-\boldsymbol{s})\times \mathrm{d}\boldsymbol{s}_1}{|\boldsymbol{s}_1-\boldsymbol{s}|^3}.
\end{equation}
Here, vortex core radius is $a_0=10^{-8}$ cm for superfluid $^4$He, circulation $ \kappa_q=h/m=9.97 \times 10^{-4}$ cm$^2$/sec is fixed by quantum constraint of the Planck constant $h$ and bare mass $m$ of a helium atom, $l_\pm$ are the lengths of the line segments connected to $\boldsymbol{s}$ after discretization, and $\boldsymbol{s}'$ and $\boldsymbol{s}''$ are unit tangent and normal vectors at point $\boldsymbol{s}$, respectively.

The computational box is a cube of size $\mathcal{L} = 0.1$ cm with periodic boundary conditions.
A tree algorithm with opening angle $\theta_{\mathrm{tree}}=0.4$ is used to speed up the evaluation of Biot--Savart integrals.
The third-order Adams--Bashforth method is used for time advance with the stepping $ \textrm{d}t=5\times 10^{-5}$ sec.
The spatial resolution is based on the cutoff scale $\delta$ which is the length of a filament segment. 
Through vortex reconnection, the size of the smallest vortex filament is cut off at scale $\delta$ \cite{Araki2002Energy}.

\begin{figure}[!htb]
	\centering
	\includegraphics[width=0.9\textwidth]{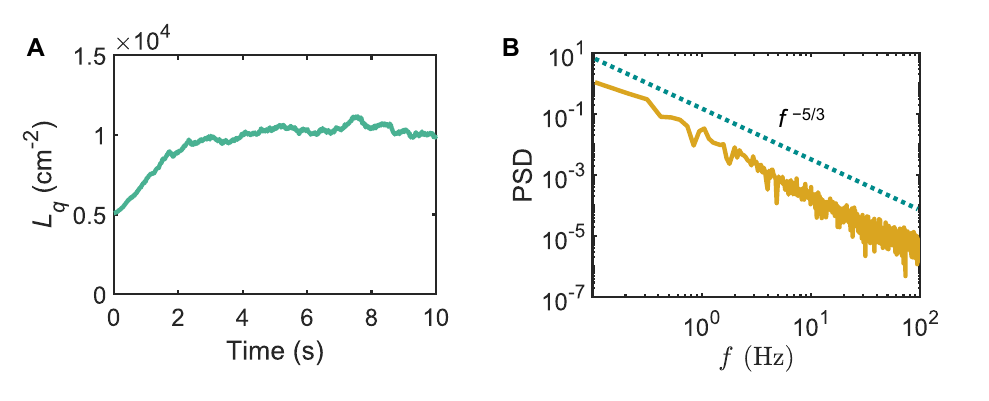}
	\caption{\textbf{Statistics of quantum turbulence with cutoff scale $\delta = 0.01$ cm.}
    (\textbf{A}) Temporal evolution of quantum vortex line density $L_q$ which is the vortex length per unit volume. After rapid initial growth, the length of the superfluid quantum vortex filament reaches a statistically stationary state.
    (\textbf{B}) Frequency spectrum (power spectral density) of the fluctuating vortex line density in the statistically stationary state. The dashed line shows the $f^{-5/3}$ scaling.}
	\label{fig:statistics_of_quantum}
\end{figure}

We performed two cases to simulate quantum vortex tangle with different degrees of entanglement by adjusting $\delta = 0.01 $ cm and $0.005$ cm (Movies S1 and S2). Both initial configurations consist of 50 vortices at random positions and with random orientations. After a rapid transition, the tangles become statistically isotropic and homogeneous (Figs.~1B and \ref{fig:statistics_of_quantum}). 
We chose the vortex tangle at $t=10$ sec in the statistically stationary stage as the skeleton of woven classical turbulence. The two VFM simulations provide two sets of vortex tangles with different complexities.

\subsection{Construction of woven classical turbulence}

\begin{figure}[!htb]
	\centering
	\includegraphics[width=0.9\textwidth]{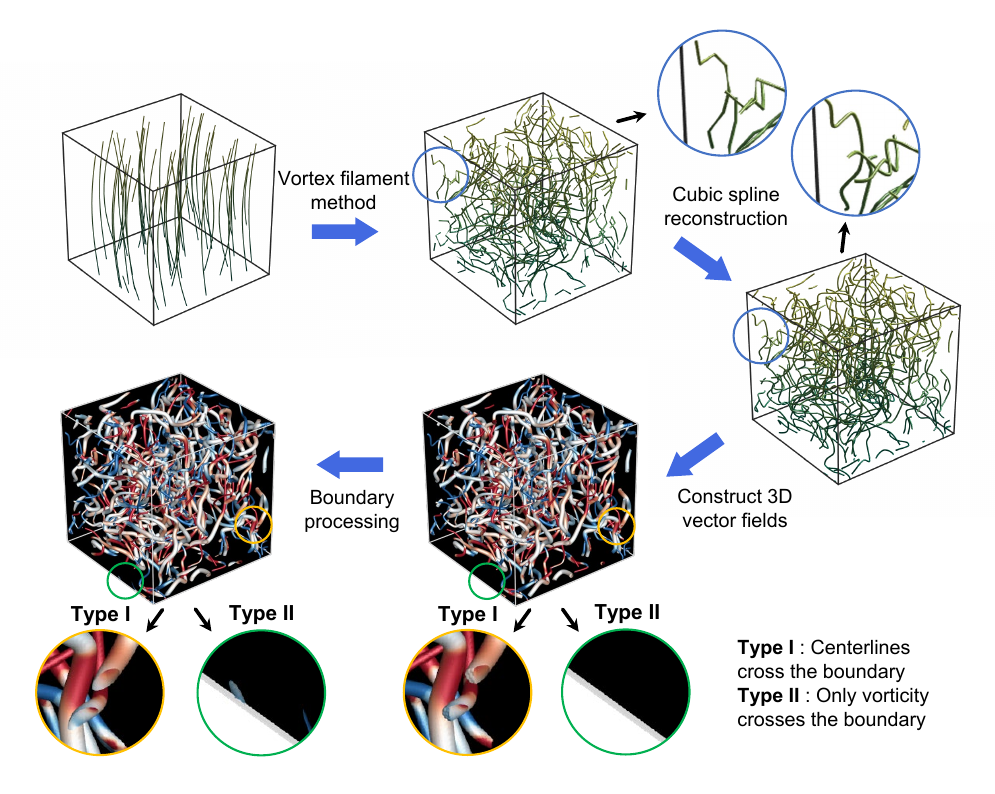}
	\caption{\textbf{Construction of woven classical turbulence with quantum skeleton.} First, the vortex filament method is used to simulate the quantum vortex tangle. Then, the control points on the filaments are transformed into a cubic spline curve defined piecewise by a 3D polynomial parametric equation. Based on these centerlines $\mathcal{C}$, the 3D vorticity field of vortex tubes can be constructed by mapping from the curvilinear coordinates centered on $\mathcal{C}$ to Cartesian coordinates. In order to satisfy the periodic boundary conditions of the vector field, we set ghost points for centerlines that cross the boundary (Type I) and consider the effect of the other $3^3-1=26$ periodic boxes around the original domain (Type II).}
	\label{fig:construction_process}
\end{figure}

In the construction of woven classical turbulence, first, we used the VFM to simulate the quantum vortex tangle.
Then, we reconstructed the control points on the filaments into cubic spline curves defined piecewise by polynomial parametric equations (Supplementary Section~\ref{subsec:cubic_spline}). Based on these centerlines $\mathcal{C}$, the 3D vorticity field for the vortex tubes was constructed in a periodic box of size $\mathcal{L} = 2\pi$ on $N^3=1024^3$ uniform grid points by mapping from the curved cylindrical coordinates centered on $\mathcal{C}$ to Cartesian coordinates (Supplementary Sections~\ref{subsec:construction_theory} and \ref{subsec:construction_numerical}).
Finally, boundary processing was performed so that the vector field satisfies periodic boundary conditions.
Two types of boundary problems need to be handled separately.
When a centerline crosses the boundary (Type I), we set ghost points for the centerline.
When the vorticity for a vortex tube with finite thickness crosses the boundary, while the centerline does not cross the boundary (Type II), we consider the effect of the other $3^3-1=26$ periodic boxes surrounding the original domain. 
The construction process is summarized in Fig.~\ref{fig:construction_process}.

\subsection{Direct numerical simulations of classical turbulence}
We performed a direct numerical simulation for solving the incompressible Navier--Stokes equations
\begin{equation}
	\frac{\partial{\boldsymbol{u}}}{\partial t}+\boldsymbol{u}\cdot \nabla\boldsymbol{u}=-\nabla p+\nu \nabla^2 \boldsymbol{u}+ \boldsymbol{f},\quad\nabla \cdot \boldsymbol{u}=0
\end{equation}
in a periodic box with the open-source pseudo-spectral code HIT3D \cite{Chumakov2008A}, where $p$ is the pressure, $\nu$ the kinematic viscosity, and $\boldsymbol{f}$ an external forcing for maintaining turbulence \cite{Machiels1997Predictability}. The forcing operates at large scales within a spherical shell of radius $|\boldsymbol{k}|\le2$ in Fourier space. The domain size is $L^3=(2\pi)^3$ with $1024^3$ grid points, and the Taylor Reynolds number is $Re_\lambda=161.4$, which matches the Reynolds number of the woven turbulence.
The turbulence statistics in this simulation are compared with those in the woven turbulence.

\section{Comparison of different synthetic turbulence methods}

We compare different synthetic turbulence methods in Table~\ref{tab:comparison_methods}.
Most existing methods can capture the scaling laws of low-order statistics of real turbulence, such as energy spectra and second-order spatial correlations. However, only our method of weaving turbulence can simultaneously reproduce low-order statistics, intermittency, and coherent vortices.

Spectral methods~\cite{Lee1992Simulation,Fung1992Kinematic} construct a turbulent field as a superposition of various Fourier modes. They can successfully achieve low-order statistical properties of the flow, such as the scaling laws of energy spectra or second-order structure functions. However, they cannot render the intermittency and high-order statistics of turbulent flows. 
Linear eddy models~\cite{Kerstein1988A} mainly estimate turbulent transport and mixing of scalars.

Fractal models include wavelet-based~\cite{Arneodo1998Random} and multiplicative methods~\cite{Eggers1992Effect,Benzi1993Extended,Juneja1994Synthetic,Biferale1998Mimicking,Scotti1999A}, which can reveal fundamental scaling properties and have been applied to model subgrid-scale stress in large eddy simulations of turbulence. These methods artificially prescribe high-order statistics but lack dynamic description. Therefore, the synthetic flows based on fractal models do not contain the basic coherent structures of turbulence.

The multiscale Lagrangian map method~\cite{Rosales2006A,Rosales2008Anomalous} can generate a 3D non-Gaussian synthetic velocity field by deforming the superposition of random-phase Fourier modes with a prescribed spectrum. It can reproduce many statistical and geometric properties. However, the main difference from real turbulence is that the regions of dominant enstrophy and low pressure are in sheet-like shapes rather than filaments.

\begin{table}[!htb]
	\centering
	\tabcolsep = 0.35cm
	\begin{tabular}{cccc}
		Methods & Low-order statistics  & Intermittency & Coherent vortices \\  \hline
		Gaussian random field  & $\times$ & $\times$ &   $\times$\\
		Spectral methods  & $\checkmark$ & $\times$ &   $\times$\\
		Linear-eddy models & $\checkmark$ & $\times$ & $\times$ \\
		Fractal models & $\checkmark$ & $\checkmark$ & $\times$ \\
		Multiscale Lagrangian map & $\checkmark$ & $\checkmark$ & sheet-like \\
		Weaving turbulence & $\checkmark$ & $\checkmark$ & customizable \\
	\end{tabular}
	\caption{\textbf{Comparison of different synthetic turbulence methods.} Low-order statistics include energy spectra and second-order spatial correlations. Intermittency is reproduced by the scaling law of high-order statistics, such as high-order velocity structure functions. The check or cross mark denotes that the feature is or is not reproduced, respectively.}
	\label{tab:comparison_methods}
\end{table}

\section{Numerical methods for weaving a turbulent field}\label{sec:construction_methods}
We detail the theoretical and numerical methods for constructing a 3D divergence-free vorticity field from a tangle of quantum vortex filaments.
Section~\ref{subsec:cubic_spline} describes how to transform a vortex filament composed of multiple spatial control points into multiple smooth parameter equations based on cubic spline polynomials.
Section~\ref{subsec:construction_theory} describes how to generate a 3D vorticity fields using vortex filaments with their centerlines in a curved cylindrical coordinate system.
Section~\ref{subsec:construction_numerical} presents a numerical algorithm for constructing a vorticity field on a Cartesian grid.
Section~\ref{subsec:boundary_processing} describes the boundary processing, and discusses the effect of different boundary processing methods on flow statistics.

\subsection{Cubic-spline reconstruction of vortex filaments}\label{subsec:cubic_spline}
In the simulation of quantum turbulence with the vortex filament method, a vortex filament is determined by a sequence of discrete control points.
To use these vortex filaments as the skeleton of vortices, we need to transform the control points of a spatial curve into an explicit parametric equation.

Under the periodic boundary conditions, all vortex filaments are either closed loops or infinitely extended with periodic repetitions. We can trace each filament from any point along itself (crossing the periodic boundary if necessary), and it will always return to the starting point, so each vortex filament can be considered closed.
For a single sequence of discrete control points  $	\widetilde{\boldsymbol{x}}_i=(x_i,y_i,z_i),~i=1,2, \ldots, N_{p}$, we introduce the discrete arc-length parameter
\begin{equation}\label{eq:sdiscrete}
	\widetilde{s}_i=\left\lbrace \begin{aligned}
		&-\left| \widetilde{\boldsymbol{x}}_{N_{p}}-\widetilde{\boldsymbol{x}}_{1} \right|,\quad i=0,\\
		&0,\quad i=1,\\
		&\sum_{j=2}^{i} \left| \widetilde{\boldsymbol{x}}_{j}-\widetilde{\boldsymbol{x}}_{j-1} \right|,\quad i=2,3, \ldots, N_{p},\\
		&\left| \widetilde{\boldsymbol{x}}_{N_{p}}-\widetilde{\boldsymbol{x}}_{1} \right|+\sum_{j=2}^{N_{p}+1} \left| \widetilde{\boldsymbol{x}}_{j}-\widetilde{\boldsymbol{x}}_{j-1} \right|,\quad i=N_{p}+1,
	\end{aligned}\right.
\end{equation}
and total discrete length $\widetilde{L}=\widetilde{s}_{N_p+1}$.
Then we obtain the normalized arc-length parameter
\begin{equation}\label{eq:zetadiscrete}
	\widetilde{\zeta}_i=\frac{2 \pi \widetilde{s}_i}{\widetilde{L}} \in[0,2 \pi],\quad i=1,2, \ldots, N_{p}+1.
\end{equation}

We use the cubic spline parametric equation
\begin{equation}\label{eq:splines}
	\boldsymbol{c}(\zeta)=\boldsymbol{\alpha}_i \zeta^3+\boldsymbol{\beta}_i \zeta^2 +\boldsymbol{\gamma}_i \zeta +\boldsymbol{\delta}_i,\quad \zeta \in [\widetilde{\zeta}_i,\widetilde{\zeta}_{i+1}),\quad i=1,2, \ldots, N_{p},
\end{equation}
to smoothly connect the control points piecewise.
In each segment $\zeta \in [\widetilde{\zeta}_i,\widetilde{\zeta}_{i+1})$, we determine four vector coefficients $[\boldsymbol{\alpha}_i, \boldsymbol{\beta}_i, \boldsymbol{\gamma}_i, \boldsymbol{\delta}_i]$ by solving the linear system 
\begin{equation}\label{eq:system}
	\begin{bmatrix}
		\boldsymbol{c}(\widetilde{\zeta}_i) \\
		\boldsymbol{c}(\widetilde{\zeta}_{i+1}) \\
		\dfrac{\textrm{d} \boldsymbol{c}}{\textrm{d} \zeta}(\widetilde{\zeta}_{i})\\
		\dfrac{\textrm{d} \boldsymbol{c}}{\textrm{d} \zeta}(\widetilde{\zeta}_{i+1})
	\end{bmatrix}=\begin{bmatrix}
		\widetilde{\zeta}_i^3 & \widetilde{\zeta}_i^2 & \widetilde{\zeta}_i & 1 \\
		\widetilde{\zeta}_{i+1}^3 & \widetilde{\zeta}_{i+1}^2 & \widetilde{\zeta}_{i+1} & 1 \\
		3 \widetilde{\zeta}_i^2 & 2 \widetilde{\zeta}_i & 1 & 0 \\
		3 \widetilde{\zeta}_{i+1}^2 & 2 \widetilde{\zeta}_{i+1} & 1 & 0
	\end{bmatrix}\begin{bmatrix}
		\boldsymbol{\alpha}_i \\
		\boldsymbol{\beta}_i \\
		\boldsymbol{\gamma}_i \\
		\boldsymbol{\delta}_i
	\end{bmatrix},
\end{equation}
where the positions
\begin{equation}\label{eq:endpositions}
	\left\lbrace \begin{aligned}
		&\boldsymbol{c}(\widetilde{\zeta}_i)=\widetilde{\boldsymbol{x}}_i\\
		&\boldsymbol{c}(\widetilde{\zeta}_{i+1})=\widetilde{\boldsymbol{x}}_{i+1}
	\end{aligned}\right.
\end{equation}
and first-order derivatives
\begin{equation}\label{eq:endderivatives}
	\left\lbrace \begin{aligned}
		&\de{\boldsymbol{c}}{\zeta}(\widetilde{\zeta}_i)=\frac{\widetilde{\boldsymbol{x}}_{i+1}-\widetilde{\boldsymbol{x}}_{i-1}}{\widetilde{\zeta}_{i+1}-\widetilde{\zeta}_{i-1}}\\
		&\de{\boldsymbol{c}}{\zeta}(\widetilde{\zeta}_{i+1})=\frac{\widetilde{\boldsymbol{x}}_{i+2}-\widetilde{\boldsymbol{x}}_{i}}{\widetilde{\zeta}_{i+2}-\widetilde{\zeta}_{i}}
	\end{aligned}\right.
\end{equation}
of the two endpoints of a segment are calculated from the positions and center differences of the control points, respectively.
Substituting Eqs.~\eqref{eq:endpositions} and \eqref{eq:endderivatives} into Eq.~\eqref{eq:system} yields
\begin{equation}\label{eq:coefficients}
	\begin{bmatrix}
		\boldsymbol{\alpha}_i \\
		\boldsymbol{\beta}_i \\
		\boldsymbol{\gamma}_i \\
		\boldsymbol{\delta}_i
	\end{bmatrix}=\begin{bmatrix}
		-\dfrac{2}{(\widetilde{\zeta}_{j}-\widetilde{\zeta}_{j+1})^3} & \dfrac{2}{(\widetilde{\zeta}_{j}-\widetilde{\zeta}_{j+1})^3} & \dfrac{1}{(\widetilde{\zeta}_{j}-\widetilde{\zeta}_{j+1})^2} & \dfrac{1}{(\widetilde{\zeta}_{j}-\widetilde{\zeta}_{j+1})^2} \\
		\dfrac{3(\widetilde{\zeta}_{j}+\widetilde{\zeta}_{j+1})}{(\widetilde{\zeta}_{j}-\widetilde{\zeta}_{j+1})^3} & -\dfrac{3(\widetilde{\zeta}_{j}+\widetilde{\zeta}_{j+1})}{(\widetilde{\zeta}_{j}-\widetilde{\zeta}_{j+1})^3} & -\dfrac{\widetilde{\zeta}_{j}+2 \widetilde{\zeta}_{j+1}}{(\widetilde{\zeta}_{j}-\widetilde{\zeta}_{j+1})^2} & -\dfrac{2 \widetilde{\zeta}_{j}+\widetilde{\zeta}_{j+1}}{(\widetilde{\zeta}_{j}-\widetilde{\zeta}_{j+1})^2} \\
		-\dfrac{6 \widetilde{\zeta}_{j} \widetilde{\zeta}_{j+1}}{(\widetilde{\zeta}_{j}-\widetilde{\zeta}_{j+1})^3} & \dfrac{6 \widetilde{\zeta}_{j} \widetilde{\zeta}_{j+1}}{(\widetilde{\zeta}_{j}-\widetilde{\zeta}_{j+1})^3} & \dfrac{\widetilde{\zeta}_{j+1}(2 \widetilde{\zeta}_{j}+\widetilde{\zeta}_{j+1})}{(\widetilde{\zeta}_{j}-\widetilde{\zeta}_{j+1})^2} & \dfrac{\widetilde{\zeta}_{j}(\widetilde{\zeta}_{j}+2 \widetilde{\zeta}_{j+1})}{(\widetilde{\zeta}_{j}-\widetilde{\zeta}_{j+1})^2} \\
		\dfrac{\widetilde{\zeta}_{j+1}^2(3 \widetilde{\zeta}_{j}-\widetilde{\zeta}_{j+1})}{(\widetilde{\zeta}_{j}-\widetilde{\zeta}_{j+1})^3} & \dfrac{\widetilde{\zeta}_{j}^2(\widetilde{\zeta}_{j}-3 \widetilde{\zeta}_{j+1})}{(\widetilde{\zeta}_{j}-\widetilde{\zeta}_{j+1})^3} & -\dfrac{\widetilde{\zeta}_{j} \widetilde{\zeta}_{j+1}^2}{(\widetilde{\zeta}_{j}-\widetilde{\zeta}_{j+1})^2} & -\dfrac{\widetilde{\zeta}_{j}^2 \widetilde{\zeta}_{j+1}}{(\widetilde{\zeta}_{j}-\widetilde{\zeta}_{j+1})^2}
	\end{bmatrix}\begin{bmatrix}
		\widetilde{\boldsymbol{x}}_i \\
		\widetilde{\boldsymbol{x}}_{i+1} \\
		\dfrac{\widetilde{\boldsymbol{x}}_{i+1}-\widetilde{\boldsymbol{x}}_{i-1}}{\widetilde{\zeta}_{i+1}-\widetilde{\zeta}_{i-1}} \\
		\dfrac{\widetilde{\boldsymbol{x}}_{i+2}-\widetilde{\boldsymbol{x}}_{i}}{\widetilde{\zeta}_{i+2}-\widetilde{\zeta}_{i}}
	\end{bmatrix}.
\end{equation}

A special treatment is needed for where two consecutive control points are on opposite sides of a boundary.
To make sure that the spline curves cross the boundary correctly and meet the periodic boundary conditions, we set ghost points on both sides of the boundary. Finally, we reconstruct all $N_{v}$ vortex filaments composed of discrete points $\widetilde{\boldsymbol{x}}_i$ into $N_{v}$ smooth cubic splines $\mathcal{C}^{(k)}:\boldsymbol{c}^{(k)}(\zeta),~k=1,2, \ldots, N_{v}$, determined by Eq.~\eqref{eq:splines} with continuous parameters $\zeta \in [0,2\pi)$ and coefficients in Eq.~\eqref{eq:coefficients}.

For any given vortex filament, the total arc-length of the spline is no less than the total discrete length of the initial sequence of control points, i.e.,
\begin{equation}\label{key}
	L=\oint \left|\de{\boldsymbol{c}(\zeta)}{\zeta}\right| d \zeta \geqslant \widetilde{L},
\end{equation}
where the equality only holds when the filament is a straight line.
Therefore, the continuous arc-length parameter
\begin{equation}\label{eq:szeta}
	s(\zeta)=\int_0^\zeta \left|\de{\boldsymbol{c}(\zeta)}{\zeta}\right| d \zeta \in[0,L)
\end{equation}
of a spline $\mathcal{C}$ is inconsistent with the discrete arc-length parameter $\widetilde{s}_i \in[0,\widetilde{L})$ of the control points on $\mathcal{C}$, and the mapping $\widetilde{s}_i \mapsto \widetilde{\zeta}_i \mapsto s(\zeta=\widetilde{\zeta}_i)$ is bijective by Eqs.~\eqref{eq:zetadiscrete} and \eqref{eq:szeta}.

\subsection{Theoretical construction of the vorticity field}\label{subsec:construction_theory}
We construct the vorticity field $ \boldsymbol{\omega} $ for multiple tube-like vortices with tunable internal structures.
These vortices are generated based on their centerlines described Eq.~\eqref{eq:splines}.
This construction method with its numerical algorithm is an extension of that in Refs.~\cite{Xiong2019Construction,Xiong2020Effects,Shen2023Role} by incorporating various simple vortex models and entanglement of multiple vortices.

\begin{figure}
	\centering
	\includegraphics[width=0.9\textwidth]{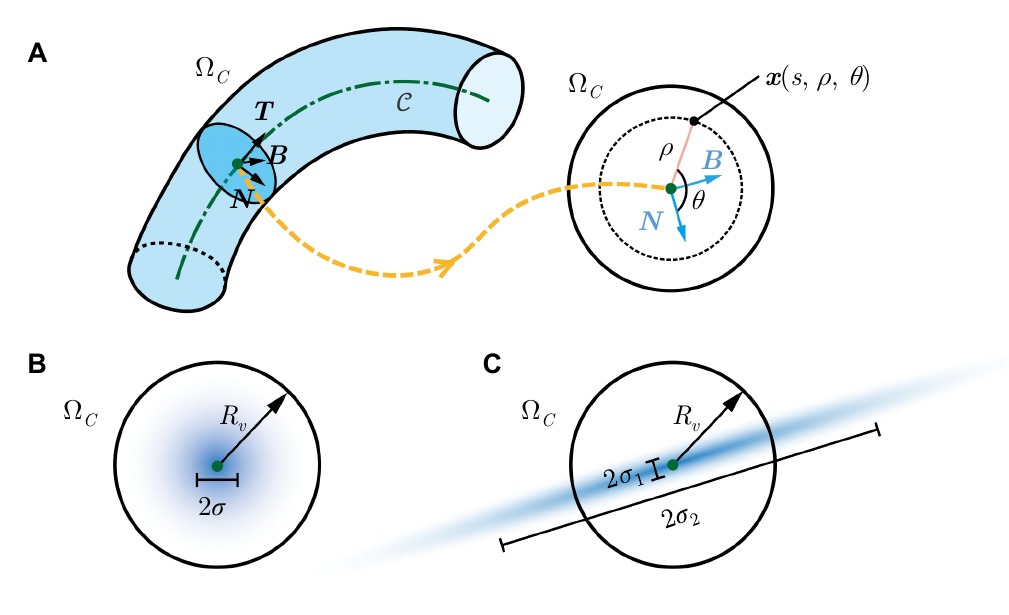}
	\caption{\textbf{Construction and coordinate system of a vortex.}
    (\textbf{A}) A segment of the tubular region $\Omega_C$, where the vorticity is constructed based on the curved cylindrical coordinates $(s,\rho,\theta)$ and the vortex centerline $\mathcal{C}$ (green dash-dotted) is described in the Frenet--Serret frame $(\boldsymbol{T}, \boldsymbol{N}, \boldsymbol{B})$.
    (\textbf{B},\textbf{C}) Cross-sectional profile of the tubular region $\Omega_C$ in the constructed vorticity field for (B) a vortex tube with thickness $2\sigma$ and (C) a vortex sheet with thickness $2\sigma_1$ and width $R_v$.}
	\label{fig:vorticity_profile}
\end{figure}

For a single vortex centerline $\mathcal{C}$, we introduce the curved cylindrical coordinate system $(s,\rho,\theta) $ \cite{Chui1995The,Xiong2020Effects} based on the Frenet--Serret frame $(\boldsymbol{T},\boldsymbol{N},\boldsymbol{B})$ in a tubular region surrounding curve $\mathcal{C}$.
The system satisfies
\begin{equation}
	\boldsymbol{x}(s,\rho,\theta)=\boldsymbol{c}(s)+\rho \cos \theta \boldsymbol{N}(s)+\rho \sin \theta \boldsymbol{B}(s)
\end{equation}
within a tubular region
\begin{equation}\label{key}
	\Omega_C=\left\{\boldsymbol{x}\left|\boldsymbol{x} \in \mathbb{R}^3, \min _s\right| \boldsymbol{x}-\boldsymbol{c}(s) \mid<R_v\right\},
\end{equation}
where $R_v$ denotes the radius of $\Omega_C$.
The Frenet--Serret formulas on $\mathcal{C}$ are
\begin{equation}\label{eq:frenet}
	\left\{\begin{aligned}
		\frac{\mathrm{d} \boldsymbol{T}}{\mathrm{d} s}&=\kappa \boldsymbol{N}, \\
		\frac{\mathrm{d} \boldsymbol{N}}{\mathrm{d} s}&=-\kappa \boldsymbol{T}+\tau \boldsymbol{B}, \\
		\frac{\mathrm{d} \boldsymbol{B}}{\mathrm{d} s}&=-\tau \boldsymbol{N},
	\end{aligned}\right.
\end{equation}
where $\boldsymbol{T} \equiv \mathrm{d} \boldsymbol{c} / \mathrm{d} s$ denotes the unit tangent, $\boldsymbol{N} \equiv(\mathrm{d} \boldsymbol{T} / \mathrm{d} s) /|\mathrm{d} \boldsymbol{T} / \mathrm{d} s|$ the unit normal, $\boldsymbol{B} \equiv\boldsymbol{T} \times \boldsymbol{N}$ the unit binormal, $\kappa$ the curvature, and $\tau$ the torsion of $\mathcal{C}$.
Based on coordinates ($s, \rho, \theta$), we specify the vorticity field
\begin{equation}\label{eq:construction1}
	\boldsymbol{\omega}(s, \rho, \theta)= \omega_s(s,\rho, \theta)\boldsymbol{e}_{s}+\omega_\rho(s,\rho,\theta)\boldsymbol{e}_{\rho}+\omega_\theta(s,\rho,\theta)\boldsymbol{e}_{\theta}
\end{equation}
of the tubular region, where the local frame is spanned by unit vectors
\begin{equation}\label{eq:localframe}
	\left\lbrace \begin{array}{l}
		\boldsymbol{e}_{s}=\boldsymbol{T}, \\
		\boldsymbol{e}_{\rho}=\cos \theta \boldsymbol{N}+\sin \theta \boldsymbol{B}, \\
		\boldsymbol{e}_{\theta}=-\sin \theta \boldsymbol{N}+\cos \theta \boldsymbol{B}.
	\end{array}\right.
\end{equation}
Figure~\ref{fig:vorticity_profile}A illustrates a segment of the tubular region $\Omega_C$ in the curved cylindrical coordinate system $(s,\rho,\theta)$.

We construct $\boldsymbol{\omega}$ based on the geometric relation between vortex lines and surfaces.
The vortex surface field (VSF) $\phi_v$ is defined to satisfy the constraint \cite{Yang2010On}
\begin{equation}\label{eq:vsf}
	\boldsymbol{\omega} \cdot \nabla \phi_v=0,
\end{equation}
so that $\boldsymbol{\omega}$ is tangent to the isosurface of $\phi_v$ everywhere except at $\boldsymbol{\omega}=0$, and every isosurface of $\phi_v$ is a vortex surface consisting of vortex lines.

For vortex tubes with the variable core size $ \sigma(s) $ and local twist rate $\eta_t(s,\phi_v) $ \cite{Shen2023Role}, the vorticity of vortex tubes with differential twist and variable thickness is specified as
\begin{equation}\label{eq:tubeconstruction}
	\boldsymbol{\omega}(s, \rho, \theta)=\Gamma f_t(s,\rho)\left[\underbrace{\boldsymbol{e}_{s}}_{\text {flux}}+\underbrace{\frac{\rho}{\sigma(s)(1-\kappa(s) \rho \cos \theta)}\frac{\mathrm{d} \sigma(s)}{\mathrm{d} s}\boldsymbol{e}_{\rho}}_{\text {tube thickness}}+\underbrace{\frac{\rho \eta_t(s, \phi_v) }{1-\kappa(s) \rho \cos \theta}\boldsymbol{e}_{\theta}}_{\text {twist}}  \right],
\end{equation}
with the circulation $\Gamma$, Gaussian kernel function
\begin{eqnarray}\label{eq:Gaussian}
	f_t(s,\rho)=\left\lbrace \begin{aligned}
		&\frac{1}{2\pi\sigma(s)^2}\exp\left[ \frac{-\rho^2}{2\sigma(s)^2}\right] ,\quad \rho \in [0,R_v),\\
		&0,\quad \rho \in [R_v,+\infty)
	\end{aligned} \right.
\end{eqnarray}
as in the Burgers vortex model, and the normalized VSF
\begin{equation}\label{eq:phiv}
	\phi_{v}(s,\rho)= 2\pi\sigma(s)^2f_t(s,\rho) \in [0,1].
\end{equation}
Here, the three terms on the right-hand side of Eq.~\eqref{eq:tubeconstruction} represent the vorticity flux, tube-thickness, and twist terms of $\bs \omega$, respectively.
The vector field constructed by Eq.~\eqref{eq:tubeconstruction} is proved to be divergence-free~\cite{Shen2023Role}.
The radius $R_v$ of the constructed tubular region $\Omega_C$ is large enough to ensure that almost all circulation is included.
The cross-sectional profile of the vortex tubes is shown in Fig.~\ref{fig:vorticity_profile}B.

For constructing sheet-like vortices, we introduce a 2D oblate normal kernel function
\begin{eqnarray}\label{eq:sheet_profile}
	f_s(s,\rho, \theta)=\left\lbrace \begin{aligned}
		&\frac{ 1}{2 \pi\sigma_1 \sigma_2} \exp \left[-\frac{1}{2}\left(\frac{\rho^2 \cos^2 \theta}{\sigma_1^2}+\frac{\rho^2 \sin^2 \theta}{\sigma_2^2}\right)\right] ,\quad \rho \in [0,R_v),\\
		&0,\quad \rho \in [R_v,+\infty)
	\end{aligned} \right.
\end{eqnarray}
with $\sigma_2\gg\sigma_1$. 
We truncate the elliptical distribution at $\rho=R_v$, so that the vorticity on a cross section is concentrated within a sheet-like region, which is centered on $\mathcal{C}$, and has thickness $2\sigma_1$ and width $2R_v$ (see Fig.~\ref{fig:vorticity_profile}C).
However, this operation reduces the total circulation, so we specify the vorticity
\begin{equation}\label{eq:sheetconstruction}
	\boldsymbol{\omega}(s, \rho, \theta)=\frac{\Gamma }{\Phi}f_s(s,\rho, \theta) \boldsymbol{e}_{s}
\end{equation}
with a compensation coefficient
\begin{equation}\label{eq:compensation}
	\Phi=\int_0^{R_v} \int_0^{2\pi}  \frac{ \rho}{2 \pi\sigma_1 \sigma_2} \exp \left[-\frac{1}{2}\left(\frac{\rho^2 \cos^2 \theta}{\sigma_1^2}+\frac{\rho^2 \sin^2 \theta}{\sigma_2^2}\right)\right] \mathrm{d} \theta \mathrm{d} \rho
\end{equation}
to make the circulation equal to $\Gamma$.

Finally, we obtain the vorticity
\begin{equation}\label{eq:addup}
	\boldsymbol{\omega}=\sum_{k=1}^{N_v} \boldsymbol{\omega}^{(k)}
\end{equation}
for a turbulent field consisting of $N_v$ intertwined vortex tubes by adding up the vorticities separately constructed from the skeleton $\mathcal{C}^{(k)},~k=1,2, \ldots, N_{v}$.

\subsection{Numerical construction of the vorticity field}\label{subsec:construction_numerical}
It is straightforward to extend the numerical algorithm in Ref.~\cite{Shen2023Role} to compute Eq.~\eqref{eq:tubeconstruction} or \eqref{eq:sheetconstruction} on a Cartesian grid with $N_x\times N_y \times N_z$ uniform grid points.
For a given parametric curve $\mc C: c(\zeta)$ with $\zeta \in\left[0, 2\pi\right)$, we divide $\mc C$ into $N_{C}$ segments by $N_{C}$ dividing points
\begin{equation}
	\boldsymbol{c}_{i}=\boldsymbol{c}\left(\zeta_{i}\right), \quad i=1,2, \ldots, N_{C},
\end{equation}
with $\zeta_{i}=(i-1) \Delta \zeta$ and $\Delta \zeta=2\pi / N_{C}$. Note that $\zeta$ is not necessary to be an arc-length parameter $s$ because of the one-to-one mapping between $\zeta$ and $s$.
Then the space in the proximity of curve $\mc C$ can be divided into $N_{C}$ subdomains
\begin{equation}\label{eq:subdomain}
	\Omega_{i}=\left\{\boldsymbol{x} \mid\left(\boldsymbol{x}-\boldsymbol{c}_{i}\right) \cdot \boldsymbol{T}_{i} \geqslant 0 \text { and }\left(\boldsymbol{x}-\boldsymbol{c}_{i+1}\right) \cdot \boldsymbol{T}_{i+1}<0\right\},
\end{equation}
with
\begin{equation}\label{eq:Ti}
	\boldsymbol{T}_{i}=\frac{\boldsymbol{c}_{i+1}-\boldsymbol{c}_{i}}{\left|\boldsymbol{c}_{i+1}-\boldsymbol{c}_{i}\right|}, \quad i=1,2, \ldots, N_{C},
\end{equation}
where subscripts $N_{C}+1$ and $1$ are equivalent.
For a given $\boldsymbol{x}$, we first use Eq.~\eqref{eq:subdomain} to determine the subdomains $\Omega_{i}$ containing $\boldsymbol{x}$. The subscripts of all the $\Omega_{i}$ containing $\boldsymbol{x}$ are denoted by a set
\begin{equation}\label{eq:Izeta}
	\widetilde{I}_{\zeta}(\boldsymbol{x})=\left\{j \mid \boldsymbol{x} \in \Omega_{j}\right\} .
\end{equation}
For each $j \in \widetilde{I}_{\zeta}(\boldsymbol{x})$, the parameter of $\mc C$ is approximated by
\begin{equation}\label{eq:conszeta}
	\widetilde{\zeta}_{j}=\frac{\zeta_{j+1}\left(\boldsymbol{x}-\boldsymbol{c}_{j}\right) \cdot \boldsymbol{T}_{j}+\zeta_{j}\left(\boldsymbol{c}_{j+1}-\boldsymbol{x}\right) \cdot \boldsymbol{T}_{j}}{\left|\boldsymbol{c}_{j+1}-\boldsymbol{c}_{j}\right|}.
\end{equation}

At $\widetilde{\boldsymbol{c}}_{j}=\boldsymbol{c}\left(\widetilde{\zeta}_{j}\right)$, we use the second-order finite difference scheme to calculate the Frenet-Serret frame
\begin{equation}\label{eq:consTNB}
	\left\lbrace \begin{aligned}
		\widetilde{\boldsymbol{T}}_{j} & =\boldsymbol{T}\left(\widetilde{\zeta}_{j}\right), \\
		\widetilde{\boldsymbol{N}}_{j} & =\boldsymbol{N}\left(\widetilde{\zeta}_{j}\right), \\
		\widetilde{\boldsymbol{B}}_{j} & =\boldsymbol{B}\left(\widetilde{\zeta}_{j}\right),
	\end{aligned}\right.
\end{equation}
as well as 
\begin{equation}\label{eq:consks}
	\left\lbrace \begin{aligned}
		&\widetilde{\kappa}_{j}=\kappa\left(\widetilde{\zeta}_{j}\right),\\
		&\widetilde{\de{\sigma}{s} }_j=\de{\sigma}{s}\left(\widetilde{\zeta}_{j}\right)
	\end{aligned}\right.
\end{equation}
in Eq.~\eqref{eq:tubeconstruction}.
In addition, the distance from $\boldsymbol{c}\left(\widetilde{\zeta}_{j}\right)$ is calculated by
\begin{equation}\label{eq:consrho}
	\widetilde{\rho}_{j}=\left|\boldsymbol{x}-\widetilde{\boldsymbol{c}}_{j}\right|,
\end{equation}
and azimuth-related functions are calculated by
\begin{equation}\label{eq:constheta}
	\left\lbrace \begin{aligned}
		\cos \widetilde{\theta}_{j}=\frac{\left(\boldsymbol{x}-\widetilde{c}_{j}\right) \cdot \widetilde{\boldsymbol{N}}_{j}}{\widetilde{\rho}_{j}}, \\
		\sin \widetilde{\theta}_{j}=\frac{\left(\boldsymbol{x}-\widetilde{c}_{j}\right) \cdot \widetilde{\boldsymbol{B}}_{j}}{\widetilde{\rho}_{j}}.
	\end{aligned}\right.
\end{equation}

We approximate Eq.~\eqref{eq:tubeconstruction} or \eqref{eq:sheetconstruction} as
\begin{equation}\label{eq:consfinal}
	\boldsymbol{\mathcal { \omega }}(\boldsymbol{x})=\sum_{j \in \widetilde{I}_{\zeta}(\boldsymbol{x})} \boldsymbol{\widetilde{\omega}}_{j},
\end{equation}
for vortex tubes
\begin{equation}\label{eq:tubenum}
	\boldsymbol{\widetilde{\omega}}_{j}= \left\lbrace \begin{aligned}
		&\Gamma f_t\left( \widetilde{\zeta}_{j},\widetilde{\rho}_{j}\right) \left[\widetilde{\boldsymbol{T}}_{j}+\frac{\widetilde{\rho}_{j}}{\sigma\left(\widetilde{\zeta}_{j}\right)\left( 1-\widetilde{\kappa}_{j} \widetilde{\rho}_{j} \cos \widetilde{\theta}_{j}\right) }\widetilde{\de{\sigma}{s} }_j\left(\cos \widetilde{\theta}_{j} \widetilde{\boldsymbol{N}}_{j}+\sin \widetilde{\theta}_{j} \widetilde{\boldsymbol{B}}_{j}\right)\right. \\
		&\qquad+\left. \frac{\widetilde{\rho}_{j} \eta_t\left( \widetilde{\zeta}_{j}, \phi_v\left( \widetilde{\zeta}_{j},\widetilde{\rho}_{j}\right) \right)  }{1-\widetilde{\kappa}_{j} \widetilde{\rho}_{j} \cos \widetilde{\theta}_{j}}\left(-\sin \widetilde{\theta}_{j} \widetilde{\boldsymbol{N}}_{j}+\cos \widetilde{\theta}_{j} \widetilde{\boldsymbol{B}}_{j}\right)  \right], \quad 1>\widetilde{\kappa}_{j} \widetilde{\rho}_{j} \cos \widetilde{\theta}_{j} \\
		&\mathbf{0}, \quad 1 \leqslant \widetilde{\kappa}_{j} \widetilde{\rho}_{j} \cos \widetilde{\theta}_{j}\end{aligned}\right.
\end{equation}
or vortex sheets
\begin{equation}\label{eq:sheetnum}
	\boldsymbol{\widetilde{\omega}}_{j}= \left\lbrace \begin{aligned}
		&\frac{\Gamma }{\Phi} f_s\left( \widetilde{\zeta}_{j},\widetilde{\rho}_{j},\widetilde{\theta}_{j}\right) \widetilde{\boldsymbol{T}}_{j}, \quad 1>\widetilde{\kappa}_{j} \widetilde{\rho}_{j} \cos \widetilde{\theta}_{j}, \\
		&\mathbf{0}, \quad 1 \leqslant \widetilde{\kappa}_{j} \widetilde{\rho}_{j} \cos \widetilde{\theta}_{j}\end{aligned}\right.
\end{equation}
with computed and given variables.
Finally, we construct $N_v$ intertwined vortex tubes by adding $N_v$ vorticity fields calculated by Eq.~\eqref{eq:consfinal} using Eq.~\eqref{eq:addup}.
The procedure for the numerical construction of $\boldsymbol{\omega}(\boldsymbol{x})$ is summarized in Algorithm~\ref{al:tubeconsturction}.

\begin{algorithm}\label{al:tubeconsturction} 
	\caption{Calculation of $\boldsymbol{\omega}(\boldsymbol{x})$ for intertwined vortex tubes/sheets}
	\KwIn{$\boldsymbol{x}$, $\widetilde{\boldsymbol{x}}_i$, $dx$, $dy$, $dz$, $ f_t(\zeta,\rho) $, $ f_s(\zeta,\rho,\theta) $, $\phi_v(\zeta,\rho)$, $\eta_t(\zeta,\phi_v)$, $ \sigma(\zeta) $, $ \sigma_1 $, $ \sigma_2 $, $ R_v $, $ \Gamma $, $N_x$, $N_y$, $N_z$, $N_v$, $N_p$, and $ N_C $}
	\KwOut{$\boldsymbol{\omega}(\boldsymbol{x})$}
	initialization\;
	
	\For{$k\leftarrow 1$ \KwTo $N_v$}
	{
		\For{$i\leftarrow 1$ \KwTo $N_p$}
		{
			Calculate $\widetilde{s}_i$ by Eq.~\eqref{eq:sdiscrete}\;
			Calculate $\widetilde{\zeta}_i$ by Eq.~\eqref{eq:zetadiscrete} \;
			Calculate $\boldsymbol{\alpha}_i, \boldsymbol{\beta}_i, \boldsymbol{\gamma}_i, \boldsymbol{\delta}_i$ by Eq.~\eqref{eq:coefficients}\;
			Obtain $\boldsymbol{c}(\zeta)$ by Eq.~\eqref{eq:splines} \;
		}
	\For{$i_x\leftarrow 1$ \KwTo $N_x$}
	{
	\For{$i_y\leftarrow 1$ \KwTo $N_y$}
	{
	\For{$i_z\leftarrow 1$ \KwTo $N_z$}
	{
		\For{$i\leftarrow 1$ \KwTo $N_C$}
		{
		Calculate $\widetilde{\boldsymbol{T}}_{i}$ by Eq.~\eqref{eq:Ti}\;
		Divide the space in the proximity of curve $\boldsymbol{c}(\zeta)$ into the subdomain $\Omega_i$ by Eq.~\eqref{eq:subdomain}\;
		Obtain $\widetilde{I}_{\zeta}$ by Eq.~\eqref{eq:Izeta} at $\bs x((i_x-1)dx,(i_y-1)dy,(i_z-1)dz)$\;
		}
		\For{$i\leftarrow 1$ \KwTo $N_C$}
		{
		\If{$i\in \widetilde{I}_{\zeta}$}
		{
		Calculate $\widetilde{\zeta}_{j}$ by Eq.~\eqref{eq:conszeta}\;
		Calculate $\widetilde{\boldsymbol{T}}_{j}, \widetilde{\boldsymbol{N}}_{j}$, and $\widetilde{\boldsymbol{B}}_{j}$ by Eq.~\eqref{eq:consTNB}\;
		Calculate $\widetilde{\kappa}_{j}$ and $\widetilde{\de{\sigma}{s} }_j$ by Eq.~\eqref{eq:consks}\;
		Calculate $\widetilde{\rho}_{j}$ by Eq.~\eqref{eq:consrho}\;
		Calculate $\cos \tilde{\theta}_{j}$ and $\sin \tilde{\theta}_{j}$ by Eq.~\eqref{eq:constheta}\;
		\If{vortex tube}
		{
			Calculate $\boldsymbol{\omega}^{(k)}(\boldsymbol{x})$ by Eqs.~\eqref{eq:consfinal} and \eqref{eq:tubenum} with computed and given variables\;
		}
		\If{vortex sheet}
		{
			Calculate $\Phi$ by Eq.~\eqref{eq:compensation}\;
			Calculate $\boldsymbol{\omega}^{(k)}(\boldsymbol{x})$ by Eqs.~\eqref{eq:consfinal} and \eqref{eq:sheetnum} with computed and given variables\;
		}
		}
		}

	}}}
	$\boldsymbol{\omega} \leftarrow \boldsymbol{\omega} + \boldsymbol{\omega}^{(k)}$
	}
\end{algorithm}

\subsection{Boundary processing}\label{subsec:boundary_processing}
We present two types of boundary problems at the bottom of Fig.~\ref{fig:construction_process}.
For centerlines that cross the boundary (Type I), we add ghost points to extend the centerlines.
For vorticity that crosses the boundary due to the tube thickness (Type II), we account for the effect of the other $3^3-1=26$ periodic boxes surrounding the computational domain.
These treatments increase the computational cost.

\begin{figure}
	\centering
	\includegraphics[width=0.8\textwidth]{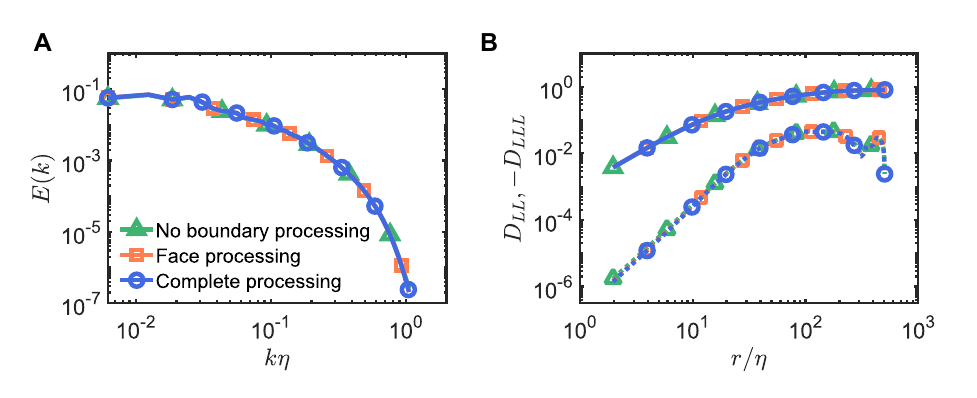}
	\caption{\textbf{Comparison of flow statistics using different boundary processing methods: no boundary processing (triangle), face processing (square), and complete processing (circle).}
(\textbf{A}) Kinetic energy spectra.
(\textbf{B}) Longitudinal second-order (solid line) and third-order (dashed line) velocity structure functions.
Woven turbulent flow fields are constructed with the core size $\sigma (s)= 0.015 \sin s$ and skeleton with $\delta = 0.01 $ cm. Here the Kolmogorov length scale is $\eta=6.22\times 10^{-3}$.}
	\label{fig:boundary_processing}
\end{figure}

To examine the influence of boundary processing on flow statistics, we compare the results from the two processing methods in Fig.~\ref{fig:boundary_processing}. We used six coplanar boxes around the flow field for face processing and all 26 boxes for complete processing. The boundary vorticity is negligible compared to the whole vorticity field, so it has a negligible effect on flow statistics.
Note that the boundary processing should be applied to the initial field to recover periodic boundary conditions in numerical simulations of flow evolution.

\section{Comparison of turbulence statistics}
We compare important statistics between woven turbulence and real classical turbulence.
First, we use the second-order structure function to estimate the mean dissipation rate in Section~\ref{sebsec:statistics}.
Then, we apply the statistical theory for homogeneous isotropic turbulence (HIT) to obtain other flow statistics and relevant length scales in Sections~\ref{sebsec:statistics} and \ref{subsec:lengthscales}.
We also compute the kinetic energy spectra and nth-order longitudinal velocity structure functions and compare them in woven and classical turbulence in Section~\ref{subsec:energy_spectrum}.
Finally, we evaluate and compare the kinetic energy flux and transfer function in Section~\ref{subsec:energy_flux}.

\subsection{Flow statistics}\label{sebsec:statistics}

\begin{figure}
	\centering
	\includegraphics[width=0.5\textwidth]{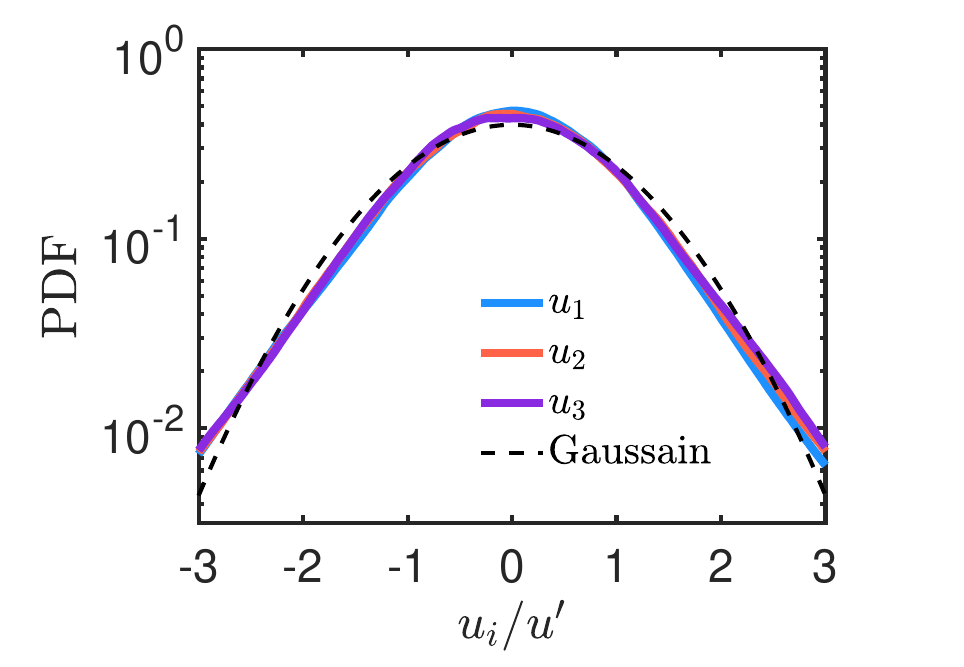}
	\caption{\textbf{Probability density functions of fluctuating velocity components.} The dashed line shows the Gaussian distribution which indicates the homogeneity and isotropy of the flow. Here, $u^{\prime}=\sqrt{\left\langle u_i u_i\right\rangle / 3}$ is the turbulent r.m.s velocity.}
	\label{fig:probability_density_functions}
\end{figure}

The velocity in woven turbulence nearly satisfies the Gaussian distribution, as shown by the probability distribution functions (PDFs) in Fig.~\ref{fig:probability_density_functions}, implying that the flow field with intertwined vortex tubes is statistically isotropic. 

In the statistical theory of HIT, the second-order longitudinal velocity structure function
\begin{equation}
	D_{LL}(r)=\left\langle\left[u_1\left(\boldsymbol{x}+\boldsymbol{e}_1 r\right)-u_1(\boldsymbol{x})\right]^2\right\rangle
\end{equation}
measures the covariance of the velocity difference between two points, $\boldsymbol{x}+\boldsymbol{e}_1 r$ and $\boldsymbol{x}$, along the same direction $\boldsymbol{e}_1$.
The celebrated Kolmogorov 1941 (K41) theory \cite{Kolmogorov1991The} suggests
\begin{equation}\label{key}
	D_{LL}(r)=C_2 \left\langle \varepsilon \right\rangle ^{2/3} r^{2/3}
\end{equation}
in the inertial range, where $\left\langle \varepsilon \right\rangle$ is the mean dissipation rate and $C_2$ is the Kolmogorov constant.
The scaling exponent of $D_{LL}$ for HIT reaches $2/3$ in the inertial range but drops below $2/3$ in the energy-containing and dissipation ranges.
Therefore, we can estimate the mean dissipation rate from the inertial range by
\begin{equation}\label{eq:estimatediss}
	\left\langle \varepsilon \right\rangle=\left[ \frac{C_2}{\max(r^{-2/3}D_{LL})}\right] ^{-3/2}
\end{equation}
where we adopt $C_2=2.2$ as a widely accepted value \cite{Su2023Simple}.

We use the mean enstrophy $\left\langle \Omega\right\rangle =\langle  \left|\boldsymbol{\omega}\right|  ^2  \rangle$ to estimate the effective kinematic viscosity
\begin{equation}\label{key}
	\nu=\frac{\left\langle \varepsilon \right\rangle}{\left\langle \Omega\right\rangle}.
\end{equation}
Finally, we calculate the Taylor microscale
\begin{equation}\label{key}
	\lambda=\sqrt{\frac{10k_t}{\left\langle \Omega\right\rangle}}
\end{equation}
and Taylor Reynolds number
\begin{equation}\label{key}
	Re_\lambda\equiv\frac{u^\prime\lambda}{\nu}=\frac{k_t}{\nu} \sqrt{\frac{20}{3\left\langle \Omega\right\rangle}},
\end{equation}
where $u^{\prime}=\sqrt{\left\langle u_i u_i\right\rangle / 3}$ is the turbulent root-mean-square (RMS) velocity and $k_t=3{u^{\prime}}^2/2$ is the turbulent kinetic energy.
Table~\ref{tab:flow_statistics} summarizes the computed values of statistics for woven turbulence as shown in Fig.~1C.

\begin{table}[]
	\centering
    \tabcolsep = 0.35cm
	\begin{tabular}{cccccc}
	Flow statistic & $k_t$   & $\left\langle \Omega\right\rangle$  & $\left\langle \varepsilon\right\rangle$  & $\nu$  & $Re_\lambda$ \\ \hline
	Value  & 3.12  & 5434.3   & 3.67  & $6.76\times 10^{-4}$    & 161.4 \\
	\end{tabular}
	\caption{Flow statistics of woven turbulence: turbulent kinetic energy $k_t$, mean enstrophy $\left\langle \Omega\right\rangle$, mean dissipation rate $\left\langle \varepsilon\right\rangle$, kinematic viscosity $\nu$, and Taylor Reynolds number $Re_\lambda$. }
	\label{tab:flow_statistics}
\end{table}

\subsection{Length scales}\label{subsec:lengthscales}
We calculate the length scales in turbulent flow fields.
The Kolmogorov length scale
\begin{equation}\label{key}
	\eta\equiv\left( \frac{\nu^3}{\left\langle \varepsilon \right\rangle}\right) ^{\frac{1}{4}}
\end{equation}
is the characteristic length scale of the smallest turbulent motion in viscous flow.
The integral length scale
\begin{equation}\label{key}
	L\equiv\frac{k_t^{3/2}}{\left\langle \varepsilon \right\rangle}
\end{equation}
characterizes the large eddies in viscous flow.
The core size
\begin{equation}
	\sigma (s)=0.015 (1+0.5\sin s)
\end{equation}
characterizes the length scale of a viscous vortex tube.

For quantum turbulence, the average distance between quantum vortices is estimated by
\begin{equation}\label{key}
	l_q = L_q^{-1/2}, 
\end{equation}
where $L_q$ is the quantum vortex line density (i.e., the vortex length per unit volume).
Since the quantum skeleton is scaled dimensionlessly into a periodic box of size $\mathcal{L}=2\pi$, the average distance between vortices in woven turbulence is
\begin{equation}\label{key}
	l = l_q\frac{\mathcal{L}}{\mathcal{L}_q},
\end{equation}
where the box size of quantum turbulence is $\mathcal{L}_q=0.1 $ cm.
Table~\ref{tab:length_scales} lists the length scales of woven turbulence.

\begin{table}[]
	\centering
    \tabcolsep = 0.25cm
	\begin{tabular}{ccccccc}
		Length scale & $\eta$   & $\sigma$  & $\lambda$  &  $l$ & $L$ & $\mathcal{L}$ \\ \hline
		Value  & $3.03\times 10^{-3}$  &  $(0.75\sim2.25)\times 10^{-2}$  &  $7.57\times 10^{-2}$ & $3.14\times 10^{-1}$  & 1.50 & $2\pi$ \\
	\end{tabular}
	\caption{Length scales of woven turbulence: Kolmogorov length scale $\eta$, core size $\sigma$, Taylor microscale $\lambda$, average distance between vortices $l$, integral length scale $L$, and box size $\mathcal{L}$. }
	\label{tab:length_scales}
\end{table}

\subsection{Energy spectrum and structure functions}\label{subsec:energy_spectrum}

\begin{figure}
	\centering
	\includegraphics[width=\textwidth]{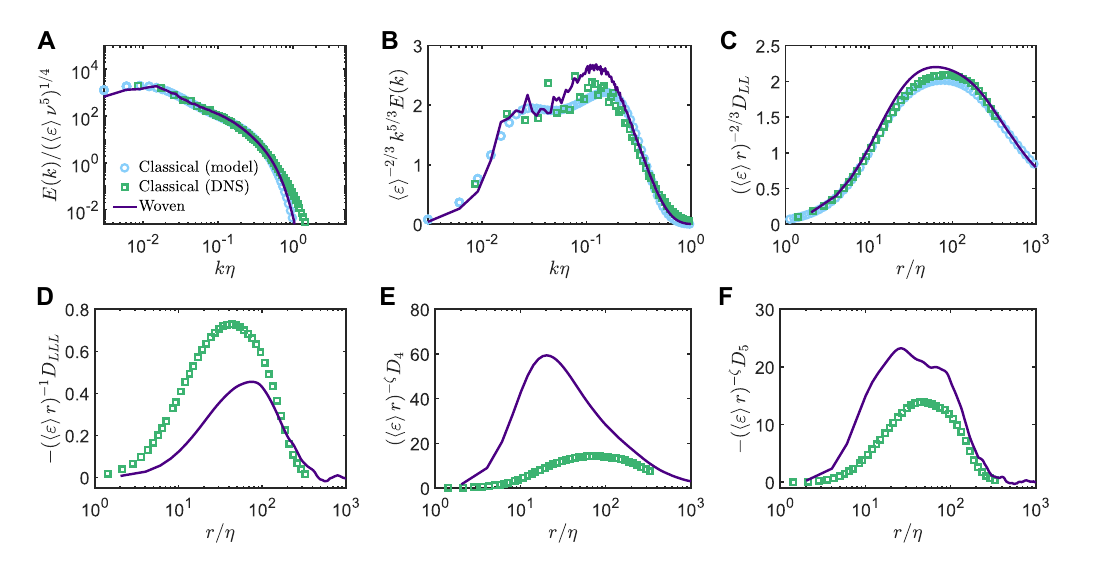}
	\caption{\textbf{Rescaled kinetic energy spectra and velocity structure functions of woven turbulence.}
    (\textbf{A}) Rescaled 3D energy spectra of classical and woven turbulence, with the Kolmogorov length scale $\eta=3.03\times 10^{-3}$, mean dissipation rate $\left\langle \varepsilon\right\rangle =3.67$, kinematic viscosity $\nu= 6.76\times10^{-4}$, and Taylor Reynolds number $Re_\lambda=161.4$.
    Lines: results for woven turbulence.
    Circles: Model results \cite{Su2023Simple} ($Re_\lambda=161.4$) for classical HIT.
    Squares: DNS results ($Re_\lambda=161.4$) of classical HIT (see Materials and Methods, Section C).
    (\textbf{B}) Rescaled compensated energy spectra of the same data as in (A).
    (\textbf{C-F}) Rescaled longitudinal nth-order velocity structure function of the same data as in (A).
    In (E) and (F), the 4th- and 5th-order structure functions are rescaled by SL94~\cite{She1994Universal} scaling exponent $\zeta(n)=n/9+2[1-(2/3)^{n/3}]$.}
	\label{fig:flow_statistics}
\end{figure}

We calculate the kinetic energy spectra and nth-order longitudinal velocity structure functions of woven turbulence and compare them with direct numerical simulation (DNS) and model results of HIT with identical flow parameters.
The 3D kinetic energy spectrum
\begin{equation}\label{eq:Ek}
	E(k)=\iint \hat{\boldsymbol{u}}^*(\boldsymbol{k}) \cdot \hat{\boldsymbol{u}}(\boldsymbol{k}) ~\mathrm{d} A(k)
\end{equation}
represents the contribution to turbulent kinetic energy by wavenumbers from $k$ to $k + \mathrm{d}k$,
where $\hat{\boldsymbol{u}}(\boldsymbol{k})$ denotes the Fourier modes of velocity field $\boldsymbol{u}$ at wavenumber $\boldsymbol{k}$, the superscript ``*'' denotes the complex conjugate, and $\mathrm{d} A(k) $ is the area of a surface element on a sphere with radius $k=\left| \boldsymbol{k}\right| $ in the Fourier space.
We plot the rescaled 3D energy spectrum of woven turbulence in Fig.~\ref{fig:flow_statistics}A and the compensated energy spectrum in Fig.~\ref{fig:flow_statistics}B.

The nth-order longitudinal velocity structure function
\begin{equation}
	D_n(r)=\left\langle\left[u_1\left(\boldsymbol{x}+\boldsymbol{e}_1 r\right)-u_1(\boldsymbol{x})\right]^n\right\rangle
\end{equation}
measures the statistical correlation of velocity differences along the direction of a unit vector $\boldsymbol{e}_1$.
The K41 theory \cite{Kolmogorov1991The} suggests
\begin{equation}\label{eq:K41_2}
	D_{LL}(r)\sim \left( \left\langle \varepsilon \right\rangle r \right) ^{2/3}
\end{equation}
and
\begin{equation}\label{eq:K41_3}
	D_{LLL}(r)\sim  \left\langle \varepsilon \right\rangle r.
\end{equation}
To account for the intermittency effects into higher-order structure functions, we adopt the She and Leveque 1994 (SL94) model \cite{She1994Universal}
\begin{equation}\label{eq:SL94}
	D_n(r) \sim  \left( \left\langle \varepsilon \right\rangle r \right) ^{\zeta(n)}
\end{equation}
with
\begin{equation}\label{eq:SL94_zeta}
	\zeta(n)=\frac{1}{9}n+2\left[ 1-\left( \frac{2}{3}\right)^{\frac{n}{3}} \right].
\end{equation}
We plot the rescaled longitudinal 2nd- to 5th-order velocity structure functions of woven turbulence in Figs.~\ref{fig:flow_statistics}C-F. Considering the intermittency corrections, the 4th- and 5th-order structure functions are rescaled with the SL94 scaling in Eq.~\eqref{eq:SL94_zeta}.

To compare statistics in woven and real classical turbulence, we performed a DNS of the incompressible Navier--Stokes equation (see Materials and Methods Section~C for more details).
Moreover, we used a simple model \cite{Su2023Simple}
\begin{equation}\label{eq:Su_model}
	D_{L L}(r)=2 {u^\prime}^2\left(\frac{r^2}{r^2+\left(15 C_2\right)^{3 / 2} \eta^2}\right)^{2 / 3-\mu / 18}\left(\frac{r^2}{r^2+\left[4 /\left(3 C_2\right)\right]^3 L^2}\right)^{1 / 3+\mu / 18}
\end{equation}
with the intermittency exponent $\mu=0.25$. 
The corresponding $E(k)$ can be calculated from Eq.~\eqref{eq:Su_model} by
\begin{equation}\label{eq:Ek2}
	E(k)=\frac{1}{2} k^3 \frac{d}{d k}\left(\frac{1}{k} \frac{d E_{11}(k)}{d k}\right),
\end{equation}
where
\begin{equation}\label{eq:E11k}
	E_{11}(k)=\frac{2 {u^\prime}^2}{\pi} \int_0^{\infty} \left( 1-\frac{1}{2 {u^\prime}^2} D_{L L}(r) \right)  \cos (k r) d r
\end{equation}
is the one-dimensional energy spectrum.

Figure~\ref{fig:flow_statistics}A shows that the energy spectrum of woven turbulence agrees well with those from the model and DNS. It not only shows the $-5/3$ scaling law in the inertial range but also exhibits the bottleneck effect of energy transfer between the inertial and dissipation ranges in Fig.~\ref{fig:flow_statistics}B.

Figures~\ref{fig:flow_statistics}B-F show that the rescaled 2nd- to 5th-order longitudinal structure functions have the same scaling law in Eq.~\eqref{eq:SL94} in the inertial range as in classical turbulence.
Note that the plateaus for the inertial range are short due to the low Reynolds number in the woven turbulence.
The negative 3rd-order structure function in Fig.~\ref{fig:flow_statistics}D suggests the negative skewness of velocity fluctuations in woven turbulence as in real classical turbulence. This consistency reveals the woven and real turbulent fields have similar fine-scale vortices, which is an essential difference from the Gaussian random field.
Moreover, the amplitudes of higher-order structure functions in woven turbulence have significant differences from those observed in DNS.
These results may be improved by using more complex fine-scale elemental vortices.

\subsection{Energy flux and transfer}\label{subsec:energy_flux}

The spectral kinetic energy flux and transfer support the classical Richardson cascade in woven turbulence in Fig.~2D.
We calculate the spectral kinetic energy flux by
\begin{equation}\label{key}
	\Pi_K(k)=\sum_{\left|\boldsymbol{k}^{\prime}\right| \leqslant k} \operatorname{Re}\left\{\boldsymbol{u}^*\left(\boldsymbol{k}^{\prime}\right) \cdot \mathcal{F}[(\boldsymbol{u} \cdot \boldsymbol{\nabla}) \boldsymbol{u}]\left(\boldsymbol{k}^{\prime}\right)\right\}.
\end{equation}
This represents the net transfer of energy from all eddies with wavenumber smaller than $k$ to those with wavenumber larger than $k$. Then, we obtain
\begin{equation}
	\Pi_K (k)=-\int_0^k T(k) d k=\int_k^{\infty} T(k) d k,
\end{equation}
where $T(k)$ is the spectral kinetic energy transfer function, which encodes the energy gain or loss at $k$.

\begin{figure}
	\centering
	\includegraphics[width=0.85\textwidth]{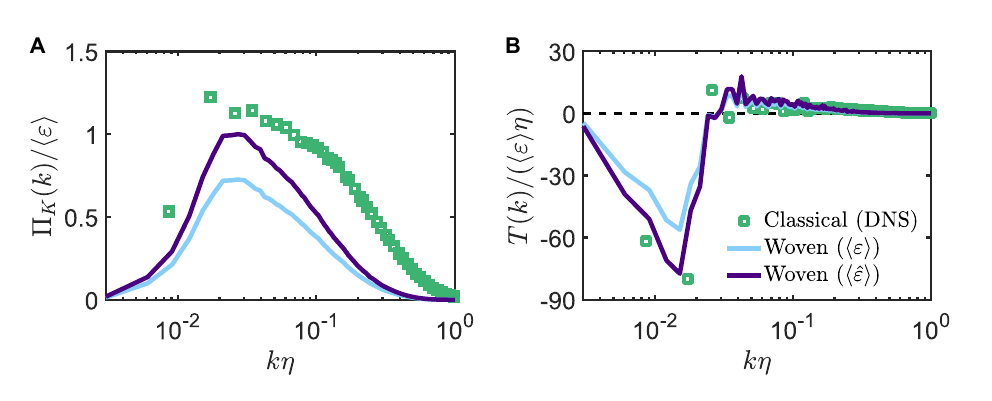}
	\caption{\textbf{Kinetic energy flux and transfer in spectral space.}
    (\textbf{A}) Rescaled spectral kinetic energy flux $\Pi_K(k)$ representing the net transfer of energy from all eddies of wavenumber less than $k$ to those of wavenumber greater than $k$. Purple lines: results rescaled using the mean dissipation rate $\left\langle \hat{\varepsilon} \right\rangle$ estimated by the energy flux; blue lines: results rescaled using $\left\langle \varepsilon \right\rangle$ estimated by the second-order structure function.
    (\textbf{B}) Rescaled spectral kinetic energy transfer function $T(k)$ representing the removal (negative) of energy from the large scales and the deposition (positive) of energy at small scales.}
	\label{fig:energy_transfer}
\end{figure}

In the inertial subrange, we expect
\begin{equation}
	\Pi_K=-\frac{\partial}{\partial t} \int_0^k E d k=\left\langle \varepsilon \right\rangle,
\end{equation}
since $\left\langle \varepsilon \right\rangle$ is the rate of loss of energy from large eddies. Thus we can also estimate the mean dissipation rate using $\langle \hat{\varepsilon} \rangle = \max(\Pi_K)=2.67$. Note that $\langle \hat{\varepsilon} \rangle\approx73\%\langle \varepsilon \rangle$ is lower than $\langle \varepsilon \rangle$ which is estimated by the second-order structure function in Eq.~\eqref{eq:estimatediss}.

In the classical turbulence theory, the kinetic energy cascade removes energy from large scales, transfers it to small scales through the inertial range, and dissipates it by viscosity at the smallest scales.
In Fig.~\ref{fig:energy_transfer}A, the energy flux increases with $k$ in the large-scale, energy containing range. It is transferred from large to small scales with a constant energy flux in the inertial range and decreases with $k$ in the small-scale, energy dissipation range.
Note that the inertial range is very short due to the low Reynolds number in the woven turbulence.
Hence, the energy transfer function is negative in the large-scale energy-containing range, positive in the small-scale dissipation range, and around zero in the inertial range. 
We compare the spectral kinetic energy flux and transfer function of woven turbulence and DNS results of classical turbulence in Fig.~\ref{fig:energy_transfer}.
The spectral energy transfer of woven turbulence agrees with that of classical turbulence.
This suggests that the viscous vortices introduced in the woven turbulence cause the shift from the quantum cascade to the classical one.

\subsection{Helicity decomposition}

The total helicity
\begin{equation}\label{eq:helicity}
	H=\iiint h \, \mathrm{d} \mathcal{V}
\end{equation}
is the volume integral of the helicity density $h = \boldsymbol{u} \cdot \boldsymbol{\omega}$.

For multiple closed vortex tubes, the helicity can be topologically decomposed into
\begin{equation}\label{eq:CW}
	H=\sum_{i \neq j} \Gamma_{i} \Gamma_{j} L_{k,ij}+\sum_{i} \Gamma_{i}^{2}(W_{r,i}+T_{w,i}),
\end{equation}
where $\Gamma_i$ is the circulation of vortex tube $i$, $L_{k,ij}$ is the Gauss linking number between vortex tubes $ i $ and $ j $, and $W_{r,i}$ and $T_{w,i}$ are the writhing and twisting numbers of vortex tube $i$, respectively \cite{Berger1984The,Scheeler2014Helicity,Shen2022Topological}.
The writhing number measures the bending or knotting of the vortex centerline $\mathcal{C}$, and the twisting number measures the twisting of the vorticity field around the centerline.

The writhing number is computed by
\begin{equation}\label{eq:Wr}
	W_r=\frac{1}{4 \pi} \oint_{\mathcal{C}} \oint_{\mathcal{C}} \frac{\left(\boldsymbol{x}-\boldsymbol{x}^{*}\right) \cdot \mathrm{d} \boldsymbol{x} \times \mathrm{d} \boldsymbol{x}^{*}}{\left|\boldsymbol{x}-\boldsymbol{x}^{*}\right|^{3}},
\end{equation}
where $\boldsymbol{x}$ and $\boldsymbol{x}^{*}$ are two points on $\mathcal{C} $.
The twisting number
\begin{equation}\label{eq:Tphi}
	T_{w}=\frac{1}{2\pi}\oint_{\mathcal{C}} \eta_t \mathrm{~d} s
\end{equation}
is defined by the local twist rate
\begin{equation}
	\eta_t \left(s\right)=\left(\boldsymbol{N}_{s} \times \boldsymbol{N}_{s}^{\prime}\right) \cdot \boldsymbol{T}
\end{equation}
of vortex lines $\mathcal{C}^{*}$ along the centerline $\mathcal{C}$,
where $\boldsymbol{N}_s$ is a radial unit vector from $\mathcal{C}$ to $\mathcal{C}^{*}$ in plane $S_C$, $\boldsymbol{N}_s^{\prime}=\mathrm{d} \boldsymbol{N}_s / \mathrm{d} s$, and $ \boldsymbol{T} $ is the unit tangent vector of $\mathcal{C}$ \cite{Shen2023Role,Chui1995The}. The Gauss linking number is given by
\begin{equation}\label{eq:Lk}
	L_{k,ij}= L_{k}(\mathcal{C}_{i},\mathcal{C}_{j}) = \frac{1}{4 \pi} \oint_{\mathcal{C}_{i}} \oint_{\mathcal{C}_{j}} \frac{\left(\boldsymbol{x}_{i}-\boldsymbol{x}_{j}\right) \cdot \mathrm{d} \boldsymbol{x}_{i} \times \mathrm{d} \boldsymbol{x}_{j}}{\left|\boldsymbol{x}_{i}-\boldsymbol{x}_{j}\right|^{3}},
\end{equation}
where $\boldsymbol{x}_i$ and $\boldsymbol{x}_j$ are points on $\mathcal{C}_i $ and $\mathcal{C}_j $, respectively.


The helical wave decomposition \cite{Waleffe1992The} can separate a flow field into purely right- or left-handed polarized parts.
In Fourier space, each wave vector has two helical wave modes for the velocity field
\begin{equation}
	\hat{\boldsymbol{u}}(\boldsymbol{k})=\hat{\boldsymbol{u}}^{+}(\boldsymbol{k})+\hat{\boldsymbol{u}}^{-}(\boldsymbol{k})=\hat{u}^{+}(\boldsymbol{k}) \boldsymbol{h}^{+}(\boldsymbol{k})+\hat{u}^{-}(\boldsymbol{k}) \boldsymbol{h}^{-}(\boldsymbol{k}),
\end{equation}
where $\boldsymbol{h}^{ \pm}$ are the eigenvectors of the curl operator that satisfy  $\mathrm{i} \boldsymbol{k} \times \boldsymbol{h}^{ \pm}= \pm|\boldsymbol{k}| \boldsymbol{h}^{ \pm}$, and $u^{ \pm}$ are the corresponding Fourier coefficients.
In physical space, the velocity field
\begin{equation}
	\boldsymbol{u}(x)=\boldsymbol{u}^R(x)+\boldsymbol{u}^L(x)+\nabla \phi
\end{equation}
can be written as a sum of complex helical waves and the gradient of a harmonic potential $\phi$, with $\boldsymbol{u}^R(\boldsymbol{x})=\int \hat{\boldsymbol{u}}^{+}(\boldsymbol{k}) \mathrm{d} \boldsymbol{k}$ and $\boldsymbol{u}^L(\boldsymbol{x})=\int \hat{\boldsymbol{u}}^{-}(\boldsymbol{k}) \mathrm{d} \boldsymbol{k}$.
Here, $\boldsymbol{u}^R$ is the projection of $\boldsymbol{u}$ onto the vector space spanned by all eigenfunctions with positive eigenvalues $(+1|\boldsymbol{k}|)$ of the curl operator, and is the right-handed component of $\boldsymbol{u}$;
$\boldsymbol{u}^L$ is a linear combination of eigenfunctions with negative eigenvalues $(-1|\boldsymbol{k}|)$, and is the left-handed component of $\boldsymbol{u}$.
For a flow that is at rest at infinity, $\nabla \phi$ is a constant vector field and can be eliminated by choosing an appropriate inertial frame.
The vorticity field can be decomposed similarly as
\begin{equation}
	\boldsymbol{\omega}(x)=	\boldsymbol{\omega}^R(x)+	\boldsymbol{\omega}^L(x),
\end{equation}
with $	\boldsymbol{\omega}^R=\nabla \times \boldsymbol{u}^R$ and $	\boldsymbol{\omega}^L=\nabla \times \boldsymbol{u}^L$.

Then, the helicity can be decomposed as
\begin{equation}
	H=H^R+H^L=\int_V 	\boldsymbol{\omega}^R \cdot \boldsymbol{u}^R \mathrm{~d} V+\int_V 	\boldsymbol{\omega}^L \cdot \boldsymbol{u}^L \mathrm{~d} V .
\end{equation}
The cross-terms $\boldsymbol{\omega}^L \cdot \boldsymbol{u}^R$ and $\boldsymbol{\omega}^R \cdot \boldsymbol{u}^L$ vanish due to the orthogonality of the eigenfunctions of the curl operator, and $H^R$ and $H^L$ are positive and negative definite, respectively.

\section{Evolution from woven turbulence}
We study the evolution of an initial woven turbulent field to examine the dynamics of woven turbulence.
We find that woven turbulence gradually evolves into the regular HIT in a decaying process.
This evolution is quantified by the turbulent kinetic energy, mean dissipation rate, and joint PDF of the second and third invariants ($Q$ and $R$) of the velocity-gradient tensor.

\subsection{DNS with the initial woven turbulence}

\begin{figure}
	\centering
	\includegraphics[width=\textwidth]{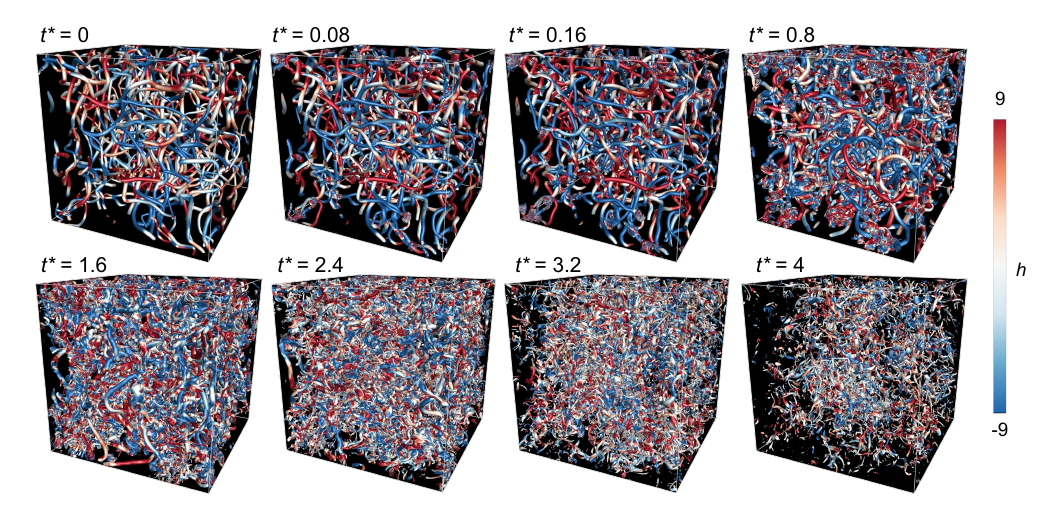}
	\caption{\textbf{Temporal evolution of vortices in decaying woven turbulence.}
    Vortex structures are visualized by the isosurface of $|\boldsymbol{\omega}|=20$ color-coded by the helicity density $h$.
    The initial vorticity field is constructed with cutoff scale $\delta = 0.01 $ cm of the quantum skeleton, and core size $\sigma= 0.03 (1+0.5\sin s)$ varies along the arc length parameter $s$. Here, the kinematic viscosity is $\nu=7.2\times 10^{-4}$ and the initial Taylor Reynolds number is $Re_\lambda=114.08$. Time is non-dimensionalized as $t^*= t/(\langle\sigma\rangle^2 / \nu)$.}
	\label{fig:decay_evolution}
\end{figure}

We first simulated a quantum vortex tangle with the cutoff scale of $0.005$ cm (Movie S2 and Materials and Methods, Section C), and chose the vortex tangle at ending time $t=10$ sec as the skeleton of woven turbulence. 
The initial vorticity field is constructed with core size $\sigma= 0.03 (1+0.5\sin s)$ varying along the arc length parameter $s$ (Fig.~\ref{fig:decay_evolution}, upper left).
This field is equivalent to the woven turbulent field in Fig.~1C in a smaller subdomain as $(1/2)^3$ of the original one, which facilitates further DNS and analysis.

We solve the 3D incompressible Navier--Stokes equations in the vorticity--velocity form using the pseudo-spectral method \cite{Meng2023Evolution,Shen2023Role} in a periodic box of size $L=2 \pi$ on uniform grid points $N^3=512^3$.
The numerical solver removes aliasing errors using the two-third truncation method with the maximum wavenumber $k_{\max} \approx N / 3$.
The time integration is treated by the explicit second-order Runge--Kutta scheme in physical space, with the adaptive time step ensuring the small enough Courant--Friedrichs--Lewy number for numerical stability and accuracy.
In this DNS, we set the kinematic viscosity to $\nu=7.2\times 10^{-4}$ to match the effective viscosity calculated from woven turbulence.

\subsection{Transition from woven turbulence to real turbulence}

Figure~\ref{fig:decay_evolution} illustrates the evolution of vortices in decaying of woven turbulence.
The entangled vortex tubes approach each other, collide and reconnect under their self-induced motion.
This interaction produces a variety of small-scale structures that fill the gaps between the initial vortex tubes.
With the intense vortex dynamics, the vortex tubes break down into smaller fragments and then dissipate by viscous effects.

\begin{figure}
	\centering
	\includegraphics[width=0.7\textwidth]{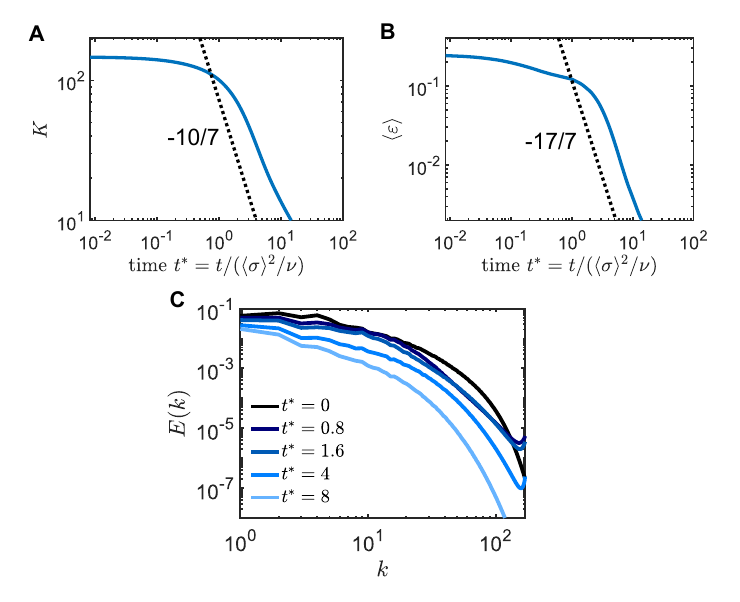}
	\caption{\textbf{Temporal evolution of flow statistics of decaying woven turbulence.}
    (\textbf{A}) Evolution of the total kinetic energy. The dotted line shows the scaling of $t^{-10/7}$. Time is non-dimensionalized as $t^*= t/(\langle\sigma\rangle^2 / \nu)$.
    (\textbf{B}) Evolution of the mean dissipation rate. The dotted line shows the scaling of $t^{-17/7}$.
    (\textbf{C}) Energy spectra at different times.}
	\label{fig:decay_statistics}
\end{figure}

In terms of the turbulent kinetic energy and mean dissipation rate in Fig.~\ref{fig:decay_statistics}, the woven turbulence first evolves into HIT in a rapid transition.
After the non-dimensionalized time $t^*= t/(\langle\sigma\rangle^2 / \nu)\approx1$, the turbulent flow decays slowly towards a quiescent state, with decay laws of $K\sim t^{-10/7}$ and $\left\langle \varepsilon \right\rangle \sim t^{-17/7}$ as in decaying HIT.
The shape of the energy spectra remains almost the same through the temporal evolution (Fig.~\ref{fig:decay_statistics}C).
This suggests that the initial woven turbulence resembles real turbulence.

\begin{figure}
	\centering
	\includegraphics[width=\textwidth]{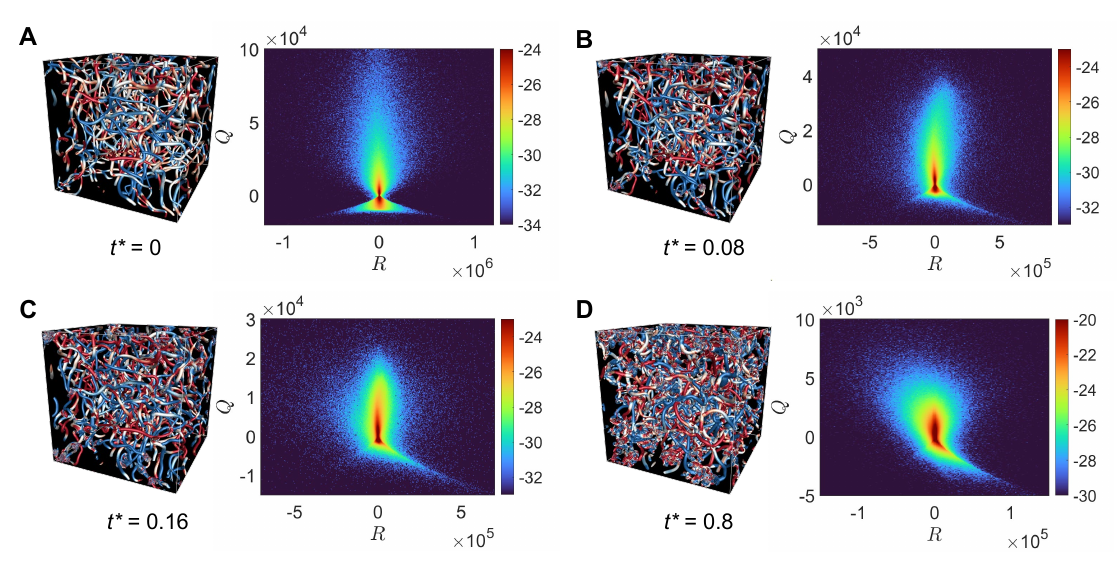}
	\caption{\textbf{Formation of the teardrop-shaped joint PDF of the second and third invariants ($Q$ and $R$) of the velocity-gradient tensor in decaying woven turbulence:} (\textbf{A}) $t^*=0$, (\textbf{B}) $t^*=0.08$, (\textbf{C}) $t^*=0.16$, and (\textbf{D}) $t^*=0.8$. Time is non-dimensionalized as $t^*= t/(\langle\sigma\rangle^2 / \nu)$. Vortex visualization is consistent with Fig.~\ref{fig:decay_evolution} as a reference.}
	\label{fig:decay_RQ}
\end{figure}

We examine how the symmetric butterfly-shaped joint PDF of $R$ and $Q$ evolves into a classic asymmetric teardrop shape during the transition from woven turbulence to real turbulence. 
Figure~\ref{fig:decay_RQ} depicts the teardrop formation.
When the vortex tubes contact, the $R$-$Q$ PDF starts to migrate along the right branch of the Vieillefosse line.
Around $t^*=0.8$, the joint PDF recovers the classic teardrop shape.
As illustrated in Figs.~\ref{fig:decay_RQ}A and D, the major difference in vortex morphology is the emergence of small-scale threads among vortex tubes during the rapid transition, while the large-scale vortex skeleton has very minor changes.
Thus, the symmetry breaking of the $R$-$Q$ PDF seems to be due to the generation of small-scale vortices.

\section{Customization of turbulence}
We describe how to customize a turbulent field by weaving different elemental vortices, which serves as a testbed for various turbulence studies.
The turbulence statistics can be manipulated by changing the core size (Section~\ref{subsec:core_size}), internal structure (Section~\ref{subsec:twist}), and cross-sectional shape (Section~\ref{subsec:sheet}) of vortex tubes, with the same quantum vortex skeleton.

\subsection{Core size}\label{subsec:core_size}
\begin{figure}
	\centering
	\includegraphics[width=0.9\textwidth]{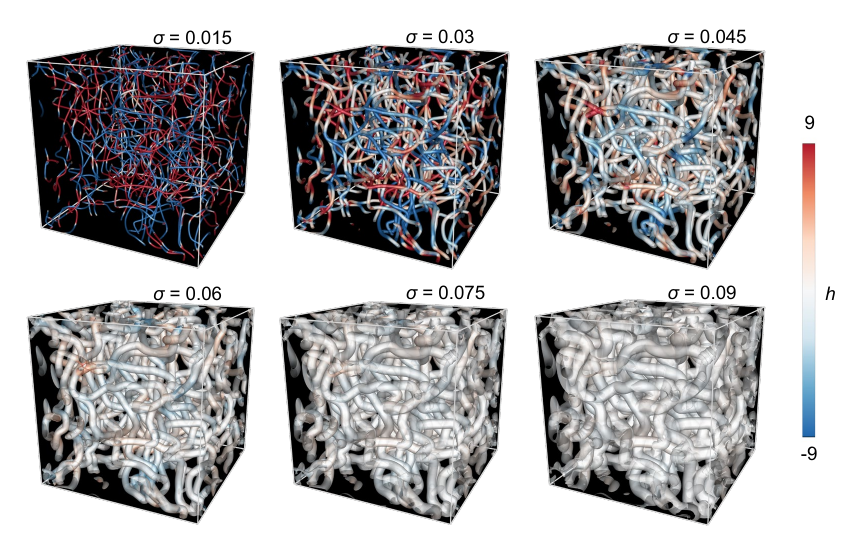}
	\caption{\textbf{Vortex visualization of woven turbulence with various core sizes.}
    The woven turbulent fields are constructed based on a quantum skeleton with the cutoff scale $\delta = 0.01 $ cm (Movie S2 and Materials and Methods, Section C). Uniform core size is set from $\sigma=0.015$ to $\sigma=0.09$. Vortex tubes are visualized by the isosurface of normalized vortex-surface field $\phi_v=0.1$, and color-coded by helicity density $h$.}
	\label{fig:core_size}
\end{figure}

The core size of the vortex tube affects the Reynolds number and the width of the inertial range (Fig.~4~D).
Using the quantum skeleton with a cutoff scale of $\delta = 0.01 $ cm (Movie S2 and Materials and Methods, Section C), we generate different woven turbulent fields with constant core sizes ranging from $\sigma=0.015$ to $\sigma=0.09$ (Fig.~\ref{fig:core_size}), corresponding to the Reynolds number from $Re = 69$ to 118.

\subsection{Internal structure}\label{subsec:twist}

The internal structure of vortices influences small-scale statistics of classical turbulence.
The vortex lines inside the vortex tube may not be parallel to the vortex centerline and may coil within the tube \cite{Xiong2019Identifying,Shen2023Role}.
The internal twist of a vortex is closely related to its helicity~\cite{Scheeler2017Complete,Shen2023Role}.
We can construct vorticity fields with highly twisted vortices by precisely controlling the local twist rate $\eta_t$.
In Fig.~4B, we created helical woven turbulence with a quantum skeleton cutoff scale of $\delta = 0.01 $ cm, and constant core size of $\sigma=0.03$, and local twist rate of $\eta_t=50$.
The right-handed internal twist wave fills the flow field with positive helicity density and right-handed chirality.

\subsection{Cross-sectional shape}\label{subsec:sheet}
\begin{figure}
	\centering
	\includegraphics[width=0.8\textwidth]{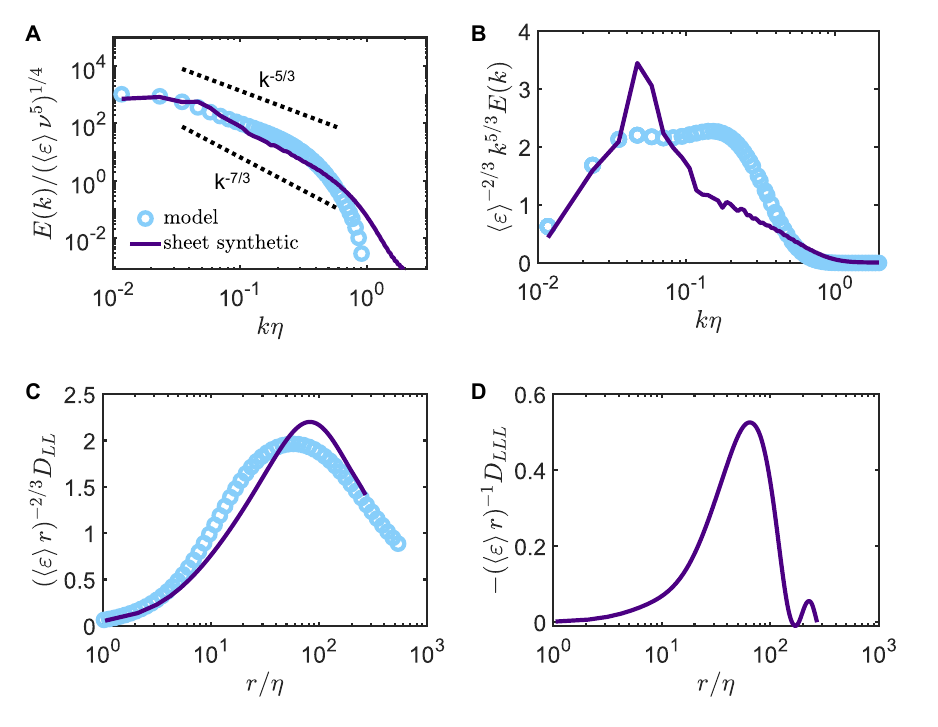}
	\caption{\textbf{ Rescaled flow statistics of vortex-sheet synthetic turbulence.}
    (\textbf{A}) Rescaled 3D energy spectra with the Kolmogorov length scale $\eta= 1.17\times 10^{-2}$, mean dissipation rate $\left\langle \varepsilon\right\rangle =4.72 \times 10^{-2}$, kinematic viscosity $\nu= 9.58\times10^{-4}$, and Taylor Reynolds number $Re_\lambda=116.4$.
    Lines: results for woven turbulence.
    Circles: model results \cite{Su2023Simple} ($Re_\lambda=116.4$) for classical HIT (see Supplementary Text, Section~\ref{subsec:energy_spectrum} for more details).
    Dotted lines: scalings of classical $k^{-5/3}$ and sheet-synthetic $k^{-7/3}$.
    (\textbf{B}) Rescaled compensated energy spectra of the same data as in (A).
    (\textbf{C-D}) Rescaled longitudinal (C) second-order and (D) third-order velocity structure function of the same data as in (A).}
	\label{fig:sheet_synthetic}
\end{figure}

With the same quantum skeleton, the cross-section of the elemental vortex can be set in various shapes.
We generate the turbulent field woven by sheet-like vortices using Eqs.~\eqref{eq:sheetconstruction}, \eqref{eq:sheet_profile}, and \eqref{eq:compensation} with sheet thickness of $\sigma_1=0.02$ and width of $R_v=\sigma_2=0.4$ (Fig.~4C).
Compared with the classical or vortex-tube woven turbulence, the statistics of the vortex-sheet woven turbulence in Fig.~\ref{fig:sheet_synthetic} show the steeper scaling law of $k^{-7/3}$ and shorter inertial range, and remain the negative third-order velocity structure function.

\clearpage

\section{Supplementary movies}
\label{sect:supplementary videos}
\subsection*{Movie S1: Evolution of quantum vortex tangle of superfluid Helium II using the vortex filament method with cutoff scale $\delta = 0.005 $ cm.}

\subsection*{Movie S2: Evolution of quantum vortex tangle of superfluid Helium II using the vortex filament method with cutoff scale $\delta = 0.01 $ cm.}

\bibliography{WeavTurb_arxiv}

\end{document}